\documentclass[epj]{svjour}
\usepackage{graphicx}
\usepackage{amsmath}
\usepackage{bm}
\usepackage{amssymb}
\usepackage{color}
\usepackage{hyperref}
\usepackage[english]{babel}
\newcommand{\aapr}{Astron. and Astrophys. Rev.}
\newcommand{\araa}{Ann. Rev. Astron. Astrophys.~}
\newcommand{\apj}{Astrophys. J.~}
\newcommand{\apjs}{Astrophys. J. Suppl.~}
\newcommand{\apjl}{Astrophys. J. Lett.~}
\newcommand{\aap}{Astron. Astrophys.~}

\newcommand{\mnras}{Mon. Not. R. Astron. Soc.~}

\newcommand{\prd}{Phys. Rev. D~}

\newcommand{\apss}{Astrophysics \& Space Science~}

\newcommand{\prl}{{Phys. Rev. Lett.}~}

\newcommand{\nat}{{Nature}~}

\newcommand{\physrep}{Phys. Rep.~}
\newcommand{\actaa}{{ACTAA}~}

\newcommand{\pasa}{Pub. Astron. Soc. Australia}
\newcommand{\pasj}{Pub. Astron. Soc. Japan}

\begin{document}
\title{GW170817---the first observed neutron star merger and its kilonova: implications for the astrophysical site of the r-process\footnotemark}

\titlerunning{GW170817---implications for the astrophysical site of the r-process}
\author{Daniel M.~Siegel\inst{1,2,3,4} 
}                     
%
%
\institute{Perimeter Institute for Theoretical Physics, Waterloo, Ontario, Canada, N2L 2Y5 \and Department of Physics, University of Guelph, Guelph, Ontario, Canada, N1G 2W1 \and Department of Physics and Columbia Astrophysics Laboratory, Columbia University, New York, NY 10027, USA  \and NASA Einstein Fellow}
\date{} 
%
\abstract{
The first neutron star (NS) merger observed by advanced LIGO and Virgo, GW170817, and its fireworks of electromagnetic counterparts across the entire electromagnetic spectrum marked the beginning of multi-messenger astronomy and astrophysics with gravitational waves. The ultraviolet, optical, and near-infrared emission was consistent with being powered by the radioactive decay of nuclei synthesized in the merger ejecta by the rapid neutron capture process (r-process). Starting from an outline of the inferred properties of this `kilonova' emission, I discuss possible astrophysical sites for r-process nucleosynthesis in NS mergers, arguing that the heaviest r-process elements synthesized in this event most likely originated in outflows from a post-merger accretion disk. I compare the inferred properties of r-process element production in GW170817 to current observational constraints on galactic heavy r-process nucleosynthesis and discuss challenges merger-only models face in explaining the r-process content of our galaxy. Based on the observational properties of GW170817 and recent theoretical progress on r-process nucleosynthesis in collapsars, I then show how GW170817 points to collapsars as the dominant source of r-process enrichment in the Milky Way. These rare core-collapse events arguably better satisfy existing constraints and overcome problems related to r-process enrichment in various environments that NS mergers face. Finally, I comment on the universality of the r-process and on how variations in light r-process elements can be obtained both in NS mergers and collapsars.
\PACS{
      {}{} 
     } 
} 
\maketitle
\section{Introduction}
\label{sec:introduction}

\renewcommand{\thefootnote}{\fnsymbol{footnote}}
\footnotetext{$\mskip-20mu ^*$ Contribution to the Topical Issue ``First joint gravitational wave and electromagnetic observations: Implications for nu- clear and particle physics'' edited by David Blaschke, Monica Colpi, Charles J. Horowitz, David Radice.}
\renewcommand{\thefootnote}{\arabic{footnote}}

In 1957, based on latest abundance data by Suess and Urey \cite{Suess1956}, Burbidge et al. \cite{Burbidge1957} and Cameron \cite{Cameron1957,Cameron57} realized that roughly half of the cosmic abundances of nuclei heavier than iron are created by rapid neutron capture onto light seed nuclei (`the r-process'). These pioneering works speculated that the high neutron fluxes required to enable neutron captures to proceed much faster than $\beta$-decays could be realized astrophysically in Type I and II supernovae. However, despite decades of intense theoretical and observational work, the cosmic origin of r-process elements has remained an enduring mystery \cite{Thielemann2011}. In particular, once considered the leading paradigm for cosmic r-process nucleosynthesis, hot outflows from a proto-neutron star in core-collapse supernovae (CCSNe) were later shown to face serious theoretical issues \cite{Qian&Woosley96,Thompson+01,Arcones+07,Roberts+12,MartinezPinedo+12} and have more recently been disfavored by several observations that point to a much rarer source of enrichment for heavy r-process elements \cite{Wallner+15,Hotokezaka+15,Ji2016} (see Sec.~\ref{sec:implications} for more details).

Lattimer \& Schramm (in 1974; \cite{Lattimer1974,Lattimer1976}) proposed that black-hole--neutron star (BH--NS) collisions give rise to significant ejection\footnote{Interestingly, the amount of ejecta material they estimated ($\sim\!0.05\,M_\odot$) matches the amount of heavy r-process material inferred from GW170817 (see Sec.~\ref{sec:EM_observations}).} of neutron-rich material whose dynamical expansion would provide a natural site for the r-process. After the discovery of the Hulse-Taylor binary pulsar \cite{Hulse1975}, Symbalisty and Schramm \cite{Symbalisty&Schramm82} noted that the inevitable collision as a result of the binary pulsar's evolution would lead to ejection of neutron-rich material, and that such NS--NS collisions might account for the entire r-process content of the galaxy. These and other pioneering works (e.g., \cite{Eichler1989,Meyer89}) were followed-up by first numerical simulations of the merger process (e.g., \cite{Davies+94,Ruffert1996b,Ruffert1997,Rosswog1999}), confirming the tidal ejection of neutron-rich material. Based on these hydrodynamic results, it was then realized with reaction network calculations that such extremely neutron-rich, tidally ejected matter with proton (electron) fraction $Y_\mathrm{e}\lesssim 0.1$ would give rise to abundances in good agreement with the solar r-process pattern for elements beyond the second r-process peak (atomic mass numbers $A>130$) \cite{Freiburghaus1999}.

Precisely six decades after the pioneering works by Burbidge et al. and Cameron, the first observed NS merger GW170817 discovered by Advanced LIGO and Virgo \cite{LIGO+17DISCOVERY} provided the first direct observation of cosmic r-process nucleosynthesis. The largest electromagnetic (EM) follow-up campaign ever conducted located the merger in the nearby SO galaxy NGC 4993 at only 40\,Mpc distance \cite{Coulter2017,Soares-Santos+17} and revealed EM counterparts across the entire EM spectrum \cite{LIGO+17CAPSTONE}, which confirmed the association of binary NS mergers with short gamma-ray bursts (GRBs). The strong thermal ultraviolet (UV), optical, and near-infrared (NIR) emission was consistent with being powered by the radioactive decay of r-process nuclei synthesized in the merger ejecta (see Sec.~\ref{sec:EM_observations}), the first unambiguous detection of a `kilonova' \cite{Li1998,Kulkarni2005,Metzger2010} (see \cite{Metzger17} for a summary on kilonova emission models).

Not only did GW170817 mark the beginning of a new era of multi-messenger astronomy and astrophysics with gravitational waves, it also provided the first observational opportunity to test theoretical knowledge about r-process nucleosynthesis and electromagnetic counterparts related to NS mergers that had been acquired over the past decades. In this paper, I shall address the quest for the cosmic origin of the heavy elements and discuss implications of GW170817 and its kilonova for the astrophysical site(s) of r-process nucleosynthesis. 

This paper is organized as follows. In Sec.~\ref{sec:EM_observations}, I briefly review the electromagnetic observations of the GW170817 kilonova and its properties, which I will then attempt to connect to specific ejecta components (astrophysical sites) in Sec.~\ref{sec:interpretation}. Section \ref{sec:implications} compares this interpretation to existing constraints on heavy r-process nucleosynthesis, which leads to a discussion of challenges that NS mergers face in explaining galactic r-process enrichment (as the sole or dominant source; Sec.~\ref{sec:challenges}). Given the inferred properties of mass ejection in GW170817 and their interpretation, I will then show how GW170817 points to collapsars as the main contributor to the galactic r-process, compare r-process nucleosynthesis in collapsars to existing constraints, and illustrate how such enrichment events overcome problems NS mergers face (Sec.~\ref{sec:collapsars}). In Sec.~\ref{sec:MHD_supernovae}, I discuss the role of MHD supernovae, another type of rare core-collapse supernovae and contender for r-process nucleosynthesis. Finally, I elaborate on how both the NS merger and collapsar scenario may account for observed variations in the light r-process nuclei while preserving the observed quasi-universal r-process pattern for heavy r-process elements (Sec.~\ref{sec:light_r_process}). Conclusions are presented in Sec.~\ref{sec:conclusion}.

\section{EM observations: kilonova properties}
\label{sec:EM_observations}

First detected 11 hours after merger \cite{Coulter2017,Soares-Santos+17}, the UV-optical-NIR transient associated with GW170817 evolved rapidly from blue colors with a spectral peak at optical wavelengths on a timescale of a day to redder (NIR) colors with a spectral peak around 1.5$\mu$m on a timescale of several days to a week \cite{Nicholl2017,Cowperthwaite2017,Kasen2017,Kasliwal2017,Drout2017,Smartt2017,Troja2017,Kilpatrick2017,Evans2017,Pian2017,McCully2017,Arcavi2017,Andreoni2017,Diaz2017,Lipunov2017,Valenti2017,Tanvir2017,Utsumi2017} (see \cite{Villar2017} for a compilation of observations from various groups and instruments). The combination of a fast rise time of $<\!11\,\mathrm{h}$, subsequent fast fading of $\gtrsim\!1\,\mathrm{mag}\,\mathrm{d}^{-1}$, the rapid color evolution from blue to red over four days, and the nearly featureless smooth optical spectra at all epochs makes this transient inconsistent with any previously observed astrophysical transient (e.g., \cite{Kilpatrick2017,Siebert2017}). 

This transient is consistent with current models for thermal emission from kilonovae \cite{Nicholl2017,Cowperthwaite2017,Kasen2017,Kasliwal2017,Drout2017,Smartt2017,Troja2017,Kilpatrick2017,Evans2017,Pian2017,McCully2017,Villar2017}. In particular, the rapid rise and fade match\-es expected kilonova peak timescales \cite{Metzger17},
\begin{equation}
	t_\mathrm{peak} \sim 1.6\,\mathrm{d} \left(\frac{M_\mathrm{ej}}{10^{-2}M_\odot}\right)^{1/2} \left(\frac{v_\mathrm{ej}}{0.1c}\right)^{-1/2} \left(\frac{\kappa}{1\,\mathrm{cm}^2\mathrm{g}^{-1}}\right)^{1/2}, \label{eq:t_peak}
\end{equation}
where $M_\mathrm{ej}$ and $v_\mathrm{ej}$ denote the ejecta mass and velocity, normalized to typical values expected for neutron star mergers (see Sec.~\ref{sec:interpretation}). The opacity $\kappa$ of the ejecta material depends on the lanthanide (and actinide) mass fraction $X_\mathrm{Lan}$. For opacities $\kappa \lesssim 1-100\,\mathrm{cm}^2\mathrm{g}^{-1}$, representative of lanthanide-free to lanthanide-rich matter \cite{Metzger2010,Kasen2013,Barnes2013,Tanaka2013}, one obtains characteristic peak timescales of kilonovae between $\sim\!1\,\mathrm{day}$ and $\sim\!1\,\mathrm{week}$, which roughly predicts the observed peak timescales of this transient (e.g., \cite{Villar2017}). Furthermore, the featureless smooth optical spectra are expected in kilonovae due to line blending resulting from the high expansion speed of the ejecta's photosphere of up to several tenths of the speed of light (\cite{Nicholl2017}; see also Sec.~\ref{sec:interpretation}). Broad features in the NIR spectra starting at $1.4-2.4\,\mathrm{d}$ post-merger have been interpreted as absorption features due to lanthanide nuclei synthesized in the merger ejecta (e.g., \cite{Kasen2013,Kasen2017,Chornock2017,Pian2017}). Finally, the bolometric lightcurve around peak light is roughly consistent with the $\propto t^{-1.3}$ decrease expected from radioactive heating of newly synthesized r-process nuclei \cite{Metzger2010,Metzger2017b}; this inference can be made due to Arnett's law, according to which the bolometric luminosity at peak light is proportional to the heating rate \cite{Arnett1979,Arnett1982}.

\subsection*{Kilonova parameters} 

The wealth of observational data from this event is consistent with a multiple-com\-po\-nent kilonova (`blue-red' kilonova). The observed light\-curves and spectra are simultaneously well fit by a two- or three-component kilonova model, in which a low-opacity, lan\-tha\-nide-free component generated the early blue emission, while a high-opacity, lanthanide-bearing component is responsible for the red emission that peaked on a time\-scale of a week \cite{Nicholl2017,Cowperthwaite2017,Chornock2017,Villar2017,Kasen2017,Kasliwal2017,McCully2017,Smartt2017,Troja2017,Pian2017,Arcavi2017,Perego+17,Coughlin2018a} (however, see \cite{Smartt2017,Tanaka2017,Waxman2018,Kawaguchi2018} and below for different approaches). The ejecta parameters for the blue component as inferred by the aforementioned papers are $v_\mathrm{ej}\approx 0.2-0.3c$, $M_\mathrm{ej}\approx 1-2\times 10^{-2}\,M_\odot$, and $X_\mathrm{Lan}\lesssim 10^{-4}$, while the red component requires $v_\mathrm{ej}\approx 0.07-0.14c$, $M_\mathrm{ej}\approx 4-6\times 10^{-2}\,M_\odot$, and $X_\mathrm{Lan} = 0.01-0.1$. For the purpose of this discussion we choose the nominal values by \cite{Cowperthwaite2017,Chornock2017,Kasen2017,Villar2017},
\begin{equation}
	\begin{array}{lll}
	v_\mathrm{ej,blue} = 0.27c, & M_\mathrm{ej,blue} = 0.016\,M_\odot, & X_\mathrm{Lan,blue}\lesssim 10^{-4},\\
	v_\mathrm{ej,red} = 0.1c, & M_\mathrm{ej,red} = 0.05\,M_\odot, & X_\mathrm{Lan,red} = 10^{-2},
	\end{array} \label{eq:KN_properties}
\end{equation}
with uncertainties defined by the above mentioned parameter ranges. While those may reasonably well cover the uncertainties in the velocities and lanthanide mass fractions, the ejecta masses likely come with larger uncertainties (see below). 

There have also been attempts to interpret the GW170817 kilonova with single-component models \cite{Smartt2017,Tanaka2017,Waxman2018}. Such models require a small but finite lanthanide mass fraction $X_\mathrm{Lan}\approx 10^{-3}$, which, in turn, requires a fine-tuned distribution in electron fraction sharply peaked around $Y_\mathrm{e}=0.25$ \cite{Lippuner2015,Tanaka+18}. This is in conflict with $Y_\mathrm{e}$ distributions from numerical simulations of the NS merger and post-merger phase, which show that ejecta components give rise to fairly broad distributions in electron fraction (\cite{Sekiguchi2016,Radice2018b,Siegel2018a,Fernandez2019,Miller2019}; see Sec.~\ref{sec:interpretation} for more details). Alternatively, such a lanthanide fraction could be obtained by mixing of lanthanide-rich and lanthanide-poor material in just the right amount if ejection of material still proceeds on the timescale $\sim\!1\,\mathrm{s}$ of the r-process (such that mixing can occur after the r-process has concluded; see also the discussion in Ref.~\cite{Metzger2018a}). However, this scenario seems unlikely, as it requires fine-tuning regarding the relative amount of ejecta material to be mixed as well as special conditions under which the ejection of material proceeds.

\subsection*{Uncertainties} 

\begin{figure}[tb]
\centering
\includegraphics[width=0.49\textwidth]{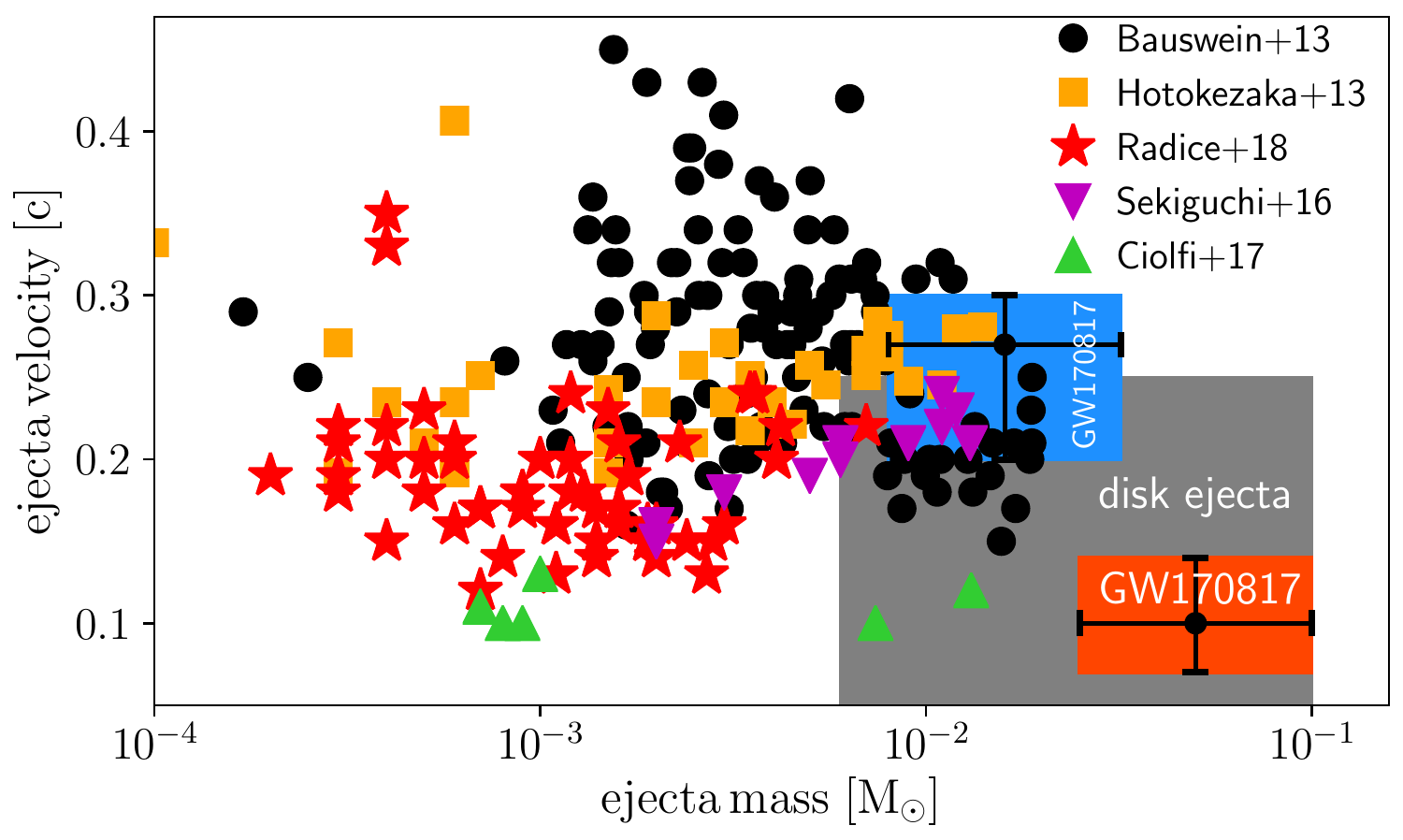}
\caption{Dynamical ejecta masses and velocities from various binary neutron star merger simulations encompassing different numerical techniques, various equations of state, binary binary mass ratios $0.65-1.0$, effects of neutrinos and magnetic fields \cite{Bauswein2013a,Hotokezaka2013b,Sekiguchi2016,Ciolfi2017a,Radice2018b}, together with the corresponding ejecta parameters inferred from the `blue' and `red' kilonova of GW170817 (see the text for details). Also shown for comparison is the parameter range for post-merger disk ejecta (see the text for details).}
\label{fig:BNS_ejecta}
\end{figure}

Systematic uncertainties in the ejecta masses not considered in the observational analyses arise from geometric and multi-dimensional effects, and uncertainties in the radioactive heating rate as well as the thermalization efficiency of the radioactive decay products. Geometric and multi-dimensional radiation transport effects may well lead to order-unity uncertainties. While those effects have been investigated to some degree \cite{Kasen+15,Wollaeger2018}, they have not yet been explored extensively in the context of GW170817 (however, see \cite{Kasen2017,Perego+17,Kawaguchi2018}). Geometric effects due to deviations from spherical symmetry have been estimated to result in mass uncertainties by a factor $\lesssim\!2$ \cite{Kasen2017}. The radioactive heating rate for `blue' (lanthanide-free) kilonova ejecta (mean electron fraction $Y_\mathrm{e}>0.25$) is uncertain by a factor of a few due to the dominance of single isotopes \cite{Lippuner2015}, its thermalization efficiency is relatively robust \cite{Metzger2010,Barnes+16}. For `red' (lanthanide-bearing) kilonova ejecta (mean electron fraction $Y_\mathrm{e}<0.25$), the radioactive heating rate is fairly robust \cite{Metzger2010,Korobkin2012,Lippuner2015}, while the thermalization efficiency is uncertain up to a factor of a few or more at late times \cite{Barnes+16,Rosswog2017a}. The latter mostly results from the sensitivity to the precise amount of translead nuclei being synthesized in the outflow (which also depends on the nuclear mass model), as translead nuclei emit a larger fraction of radioactive energy through $\alpha$-decay and fission, which thermalizes more efficiently than energy from $\beta$-decays \cite{Barnes+16}. Uncertainties in the total ejecta mass of GW170817 considering thermalization efficiency, composition, and mass model are found to be of a factor $\lesssim\!2$ \cite{Wu2019a}. Overall, this motivates at least a factor of two uncertainty in the ejecta masses of the individual components, which we shall adopt for the discussion here (see below and Figs.~\ref{fig:BNS_ejecta} and \ref{fig:BNS_rate_constraints}).



\section{Interpretation of the GW170817 kilonova}
\label{sec:interpretation}

In the light of the kilonova parameters discussed above, we turn to the question of what these observations imply for the astrophysical site of the r-process in NS mergers. We shall focus here on binary NS mergers, although the possibility of a BH--NS merger is also briefly discussed below. 

Numerical simulations of the binary NS merger and post-merger phase have revealed several types of neutron-rich ejecta over the last two decades, which we shall briefly discuss in the context of GW170817: dynamical ejecta including tidal and shock-heated components \cite{Ruffert1997,Rosswog1999,Oechslin2007,Hotokezaka2013a}, neutrino-driven and magnetically driven winds from a (meta-)stable remnant NS \cite{Dessart2009,Siegel2014a,Ciolfi2017a,Ciolfi2019}, and outflows from a post-merger neu\-trino-cooled accretion disk \cite{Fernandez2013,Just2015a,Siegel2017a}. In a typical NS mer\-ger event, all of these processes are at play to some extend, giving rise to ejecta material with different properties (amount of ejected material, composition, velocities); therefore, in principle, kilonovae with multiple components are naturally expected, which further motivates the multi-component interpretation of the GW170817 kilonova discussed above.

\subsection*{Dynamical ejecta}

During the merger process, on time\-scales of $\sim\!
\mathrm{ms}$, tidal forces tear off matter from the surfaces of the approaching neutron stars, giving rise to tidal debris streams with ejecta material mostly concentrated in the equatorial plane \cite{Hotokezaka2013a,Sekiguchi2015,Radice2018b}. Such tidal debris is composed of unprocessed, cold NS material with very low electron (proton) fraction $Y_\mathrm{e}<0.1$, which results in a non-negligible lanthanide mass fraction $X_\mathrm{lan}$ after undergoing the r-process \cite{Lippuner2015}. As the NSs collide kinetic energy is converted into thermal energy through shock heating at the contact interface between the two stars. Such shock heating, aided by pulsations of the newly-formed double-core structure, squeezes additional material into polar (low-density) regions, giving rise to a spherical ejecta structure \cite{Bauswein2013a,Hotokezaka2013a,Sekiguchi2015,Radice2018b}. The shock-heated component generally extends to higher velocities \cite{Kyutoku2014,Hotokezaka2013b,Bauswein2013a,Sekiguchi2016} and some material may overtake the tidal component ejected earlier; this can lead to interaction and partial reprocessing of the tidal ejecta by weak reactions that slightly raise the electron fraction of the affected tidal material to $Y_\mathrm{e}\approx 0.1$ \cite{Radice2018b}. Dynamical ejecta is typically characterized by ejecta masses $M_\mathrm{ej}=10^{-4}-10^{-2}\,M_\odot$ and average velocities on the order of the NS escape speed, $0.15-0.3\,c$ \cite{Bauswein2013a,Hotokezaka2013b,Dietrich2015a,Sekiguchi2015,Sekiguchi2016,Kastaun2016a,Ciolfi2017a,Radice2018b} (see Fig.~\ref{fig:BNS_ejecta} for a compilation of results from recent binary NS merger simulations). A small fraction of the shock-heated ejecta (typically $\sim 10^{-6}\,M_\odot$) is accelerated to $\gtrsim\!0.6\,c$, which can produce a bright radio transient on timescales of weeks to months or years after merger \cite{Hotokezaka2013b,Hotokezaka2018b,Radice2018b} and give rise to a freeze-out of free neutrons that could power a `neutron precursor' (blue thermal emission at timescales of hours after merger) \cite{Metzger+15}. The shock-heated ejecta is composed of hot material that has been reprocessed by weak interactions to typical average electron fractions above the critical threshold $Y_\mathrm{e}=0.25$ \cite{Lippuner2015} for lanthanide production, and is sensitive to the details of neutrino transport \cite{Radice2016,Sekiguchi2015,Sekiguchi2016,Radice2018b}. 

The overall r-process nucleosynthesis signature of dynamical ejecta and the question of whether it gives rise to a fast blue (lanthanide-free) or fast red (lanthanide-bearing) kilonova depends on the relative mass contribution of the tidal and shock-heated components, which, in turn, mostly depends on the binary mass ratio and the equation of state (EOS). For a given EOS, tuning the binary mass ratio $q=M_2/M_1< 1$ away from unity generally enhances the tidal torque on the lighter companion and thus leads to increased tidal ejecta, while reducing the shock-heated component (as the lighter companion becomes tidally elongated it seeks to avoid the collision) \cite{Hotokezaka2013b,Bauswein2013a,Dietrich2015a,Lehner2016,Sekiguchi2016}; this can change the mean $Y_\mathrm{e}$ from $>\!0.25$ to $<\!0.25$ (see, e.g., \cite{Sekiguchi2016}) and thus the nature of the kilonova component from blue to red. For a given binary mass ratio, changing the EOS from stiff (large NS radii) to soft (small NS radii) enhances the shock-heated component while reducing the tidal component \cite{Hotokezaka2013b,Bauswein2013a,Dietrich2015a,Palenzuela2015,Sekiguchi2016}. This is because tidal forces are smaller for less extended objects and NSs with smaller radii approach even closer and thus collide with higher velocities, thereby enhancing the shock power and, as a consequence, the shock-heated ejecta mass. Finally, the observational detection as a blue vs. red kilonova also depends on the geometry (viewing angle), as, e.g., tidal ejecta with its higher opacity can block blue emission from the interior on timescales of interest in equatorial directions. 

Figure~\ref{fig:BNS_ejecta} shows a compilation of dynamical ejecta properties from recent binary NS merger simulations in comparison to the GW170817 kilonova properties discussed above. The simulations shown here include different numerical techniques (smoothed-particle hydrodynamics with approximate treatment of general relativity \cite{Bauswein2013a} and several grid-based numerical relativity codes \cite{Hotokezaka2013b,Sekiguchi2016,Ciolfi2017a,Radice2018b}), various different EOS (including realistic EOS with full temperature and composition dependence), different binary mass ratios $q=0.65-1.0$. Some simulations additionally include weak interactions with different approximations to neutrino transport \cite{Sekiguchi2016,Radice2018b}, or effects of magnetic fields \cite{Ciolfi2017a}. The ejecta in some of the latter simulations contain contributions from early epochs of either neutrino-driven or magnetically driven winds (see below), which tend to decrease the overall mean ejecta velocity.

The blue kilonova properties are in tension (or at least only marginally consistent) with dynamical ejecta parameters (cf.~Fig.~\ref{fig:BNS_ejecta}). The GW170817 blue ejecta mass scale of $\sim\!10^{-2}\,M_\odot$ can be reached by nearly-equal mass binaries and soft EOSs with very small NS radii ($<\!11-12\,\mathrm{km}$) only in very few singular cases \cite{Hotokezaka2013b,Bauswein2013a,Dietrich2015a,Sekiguchi2016,Ciolfi2017a}; however, see \cite{Radice2018b} who do not find such high ejecta masses for a similar setup (e.g., SFHo with $1.35\,M_\odot$ NSs). Additionally, if one accepts the need for a massive post-merger accretion disk in order to explain the overall ejecta mass of GW170817 (see below), some EOS under which such high shock-heated ejecta masses are obtained (e.g., APR4, \cite{Hotokezaka2013b,Ciolfi2017a}) are disfavored \cite{Radice2018a}. This mass scale is instead more easily reached by a significant tidal component in asymmetric binaries with binary mass ratio significantly different from 1.0 \cite{Hotokezaka2013b,Bauswein2013a,Sekiguchi2016}, which, however, would turn the kilonova red (note, e.g., the drop in mean $Y_\mathrm{e}$ from 0.3 (SFHo 1.35-1.35) to 0.16 (SFHo 1.25-1.45) in \cite{Sekiguchi2016}).

The red kilonova properties are clearly in tension with dynamical ejecta parameters, even with generous error bars on the inferred ejecta mass (cf.~Fig.~\ref{fig:BNS_ejecta}). While similarly high ejecta masses have been obtained for extreme mass ratios and peculiar NS masses \cite{Dietrich2015a}, such ejecta being of dynamical origin would be too fast and therefore inconsistent with the low velocity $\approx\!0.1c$ inferred for the red kilonova component.

\emph{Effects of spin.} For NS with initial moderate dimensionless (but astrophysically large\footnote{The largest NS spins among the binary neutron star systems known to merge within a Hubble time are $\chi\lesssim 0.04$ after accounting for spin-down until merger \cite{Burgay03}.}) spin $\chi\approx 0.1$ aligned with the orbital angular momentum, the dynamical ejecta masses typically decrease for equal-mass binaries with respect to the irrotational case, while moderately enhancing it for spins antialigned with the orbital angular momentum \cite{Kastaun2015a,Dietrich2017a,East2019,Most2019}. Dynamical mass ejection is enhanced for cases in which tidal ejecta dominate, i.e., in asymmetric binaries \cite{Dietrich2017a,Dietrich2018a}. In the latter case, the enhancement in ejecta mass is only very moderate, but can reach up to a factor of 1.7 in special cases \cite{Dietrich2017a}. However, this only enhances the amount of fast lanthanide-rich (red) tidal ejecta, which does not help to reconcile dynamical ejecta properties neither with the red nor blue kilonova parameters.

\emph{Effects of eccentricity.} Finally, we note that eccentric mergers can enhance the dynamical ejecta masses significantly up to levels of $\sim\!10^{-2}\,M_\odot$ \cite{Gold2012,Radice2016,East2016,Chaurasia2018}, with typically decreasing ejecta masses as the eccentricity increases \cite{East2016,Chaurasia2018}. Such eccentric binaries could result from dynamical capture mergers in globular clusters as well as in other dense stellar systems, in which neutron stars may have large spins. NS spins in eccentric mergers have a moderate to strong effect on mass ejection and can have opposite effects on the total ejecta mass depending on eccentricity and spin orientation \cite{East2016,Chaurasia2018}. The dynamical ejecta are dominated by tidal ejecta \cite{Chaurasia2018}, and thus again lead to fast red transients, in conflict with the GW170817 kilonova. Furthermore, the GW signal of GW170817 \cite{LIGO+17DISCOVERY} is consistent with a quasi-circular binary, without indications of significant eccentricity.

In conclusion, it appears likely that a moderate amount of shock-heated ejecta ($\mathrm{few}\times 10^{-3}\,M_\odot$) contributed to the blue kilonova emission, but that a possibly dominant amount of blue ejecta originated in winds from the remnant NS (see below). The presence of such a strong shock-heated component would favor a nearly equal-mass merger and a soft EOS with small NS radius $<11-12\,\mathrm{km}$, which has important consequences for the EOS and nuclear phys\-ics \cite{Ozel+16}. The tidal ejecta component was probably of too small mass to be observed as an independent fast red kilonova component, which is consistent with the requirement of a nearly equal-mass merger where this component is typically sub-dominant.

\subsection*{Winds from a remnant NS} 

If the total mass $M_1+M_2$ of the binary does not exceed a certain threshold mass, which is typically between 30\% to 60\% higher than the maximum mass of non-rotating NSs (depending on the compactness of the non-rotating maximum-mass configuration) \cite{Hotokezaka2011,Bauswein2013b}, the NS merger does not lead to a prompt collapse to a BH, but instead results in a hypermassive or supramassive NS that is temporarily stable to collapse due to rapid (differential) rotation and thermal effects \cite{Shapiro2000,Duez2006a,Siegel2013,Kaplan2014}. Winds driven off the surface of such a remnant on timescales of tens to hundreds of ms represent an additional source of ejecta material. Such winds can be driven by neutrinos from the shock-heated hot ($\sim\!10\,\mathrm{MeV}$) interior of the remnant by reabsorption in a ``gain'' layer at the surface of the remnant via charge-current interactions in a way similar to neutrino-driven (thermal) winds in ordinary core-collapse supernovae \cite{Dessart2009}. Such winds give rise to initial mass loss rates of $\dot{M}\lesssim 10^{-3} M_\odot\,\mathrm{s}^{-1}$ \cite{Dessart2009} and have meanwhile been observed in dynamical merger simulations \cite{Sekiguchi2015,Sekiguchi2016,Radice2016,Radice2018b}. Magnetic field amplification in the interior of the remnant due to differential rotation can also give rise to winds launched from the surface \cite{Siegel2014a,Ciolfi2017a,Ciolfi2019}, with typical initial mass-loss rates of $\dot{M}\sim 10^{-3}-10^{-2}\,M_\odot\,\mathrm{s}^{-1}$ \cite{Siegel2014a}. The total mass loss through both types of winds is then determined by the minimum of the remnant lifetime (if metastable) and the cooling timescale or the timescale for removal of differential rotation (the latter are both $\sim\!1\,\mathrm{s}$). These winds are typically slow ($\lesssim 0.1c$) and have a mean electron fraction $\bar{Y}_\mathrm{e}>0.25$, as the ejecta material is exposed to strong neutrino emission that seeks to raise the electron fraction to an equilibrium value $Y_\mathrm{e}\approx0.5$ \cite{Qian&Woosley96}. Ref.~\cite{Metzger2018a} argues that the combination of the two processes, a neutrino-heated, magnetically-accelerated wind, could lead to substantially increased mass loss rates and ejecta velocities. For realistic lifetimes of $0.2-2\,\mathrm{s}$ of the remnant, magnetic field strengths of $B\sim 1-3\times 10^{14}\,\mathrm{G}$ and typical rotation periods of $\approx\!0.8\,\mathrm{ms}$, this model simultaneously predicts velocities $\approx\!0.3c$, ejecta masses $\approx\!0.02\,M_\odot$, and composition $\bar{Y}_\mathrm{e}\approx 0.25-0.4$ consistent with the blue kilonova component of GW170817 \cite{Metzger2018a}. Further studies are required to explore this mechanism in more detail. 

Another (perhaps more speculative) source of mass ejection is due to viscous rearrangement of the merger remnant \cite{Fujibayashi+18,Radice2018c}. However, the physical source of viscosity assumed in these studies remains unclear, as the interior of such remnants are stable against the magnetorotational instability, which is usually invoked to give rise to turbulence and effective viscosity in these systems.

\subsection*{Outflows from a post-merger accretion disk} 

After merger, a significant amount of material quickly circularizes around the remnant to form a neutrino-cooled accretion disk \cite{Shibata2006a,Fernandez2013,Just2015a,Siegel2017a,Siegel2018a,Fernandez2019,Fujibayashi+18,Miller2019}. While initial two-dimensional simulations in Newtonian gravity with an $\alpha$-viscosity prescription to parametrize angular momentum transport found comparatively small ejecta masses, more recent simulations in full 3D and general relativity including magnetic fields and self-consistent MHD turbulence as well as neutrino cooling find that $30\%-40\%$ of the original disk mass is ejected into unbound outflows \cite{Siegel2017a,Siegel2018a,Fernandez2019}. These outflows are launched as thermal winds from a hot disk corona, which is the result of an imbalance between viscous heating from magnetohydrodynamic turbulence and cooling via neutrino emission at higher latitudes off the disk midplane \cite{Siegel2017a,Siegel2018a}. Further acceleration due to $\alpha$-particle formation leads to an average outflow speed of $\approx 0.1c$ \cite{Siegel2017a,Siegel2018a}, with the bulk material distributed between $\approx(0.01-0.25)c$ \cite{Siegel2018a,Fernandez2019}. The composition of the outflows is controlled by a self-regulation mechanism based on electron degeneracy in the inner part of the disk, which keeps the mean electron fraction of the outflows $\bar{Y}_\mathrm{e}\lesssim 0.2$ well below the critical threshold for lanthanide and actinide production on timescales of interest \cite{Siegel2017a,Siegel2018a}. Detailed nucleosynthesis calculations show that the production of the full range of solar r-process abundances and abundances in metal-poor stars in the halo of the Milky Way from the first to the third peak can be explained \cite{Siegel2018a,Wu2016,Miller2019}.

Large enough disk masses to explain the red kilonova component of GW170817 require that the merger event did not result in a prompt collapse to a BH, but instead formed at least a temporarily-stable hypermassive NS \cite{Radice2018a,Radice2018b}. Excluding prompt collapse, recent simulations using various different EOS obtain typical disk masses of $0.1\,M_\odot$, with a spread between $0.02-0.24\,M_\odot$ \cite{Radice2018b}. This then translates into disk outflows of $M_\mathrm{ej}=6\times 10^{-3}-0.1\,M_\odot$ with a typical value of $M_\mathrm{ej}\approx 4\times 10^{-2}\,M_\odot$, consistent with the red kilonova of GW170817. The lifetime of the remnant must not be too long, however, as strong neutrino emission from the remnant raises the electron fraction of the disk outflows and can turn those outflows blue for lifetimes $\gtrsim\!100\,\mathrm{ms}$ \cite{Metzger2014c,Lippuner2017a,Fahlman2018}. Early non-axisymmetries in the  remnant structure may even enhance such blue outflows as a result of inducing spiral waves in the surrounding accretion disk \cite{Nedora2019}. It remains uncertain, however, whether such an additional source of viscous ejecta would operate in the presence of MHD-driven disk outflows.

\begin{figure}[tb]
\centering
\includegraphics[width=0.49\textwidth]{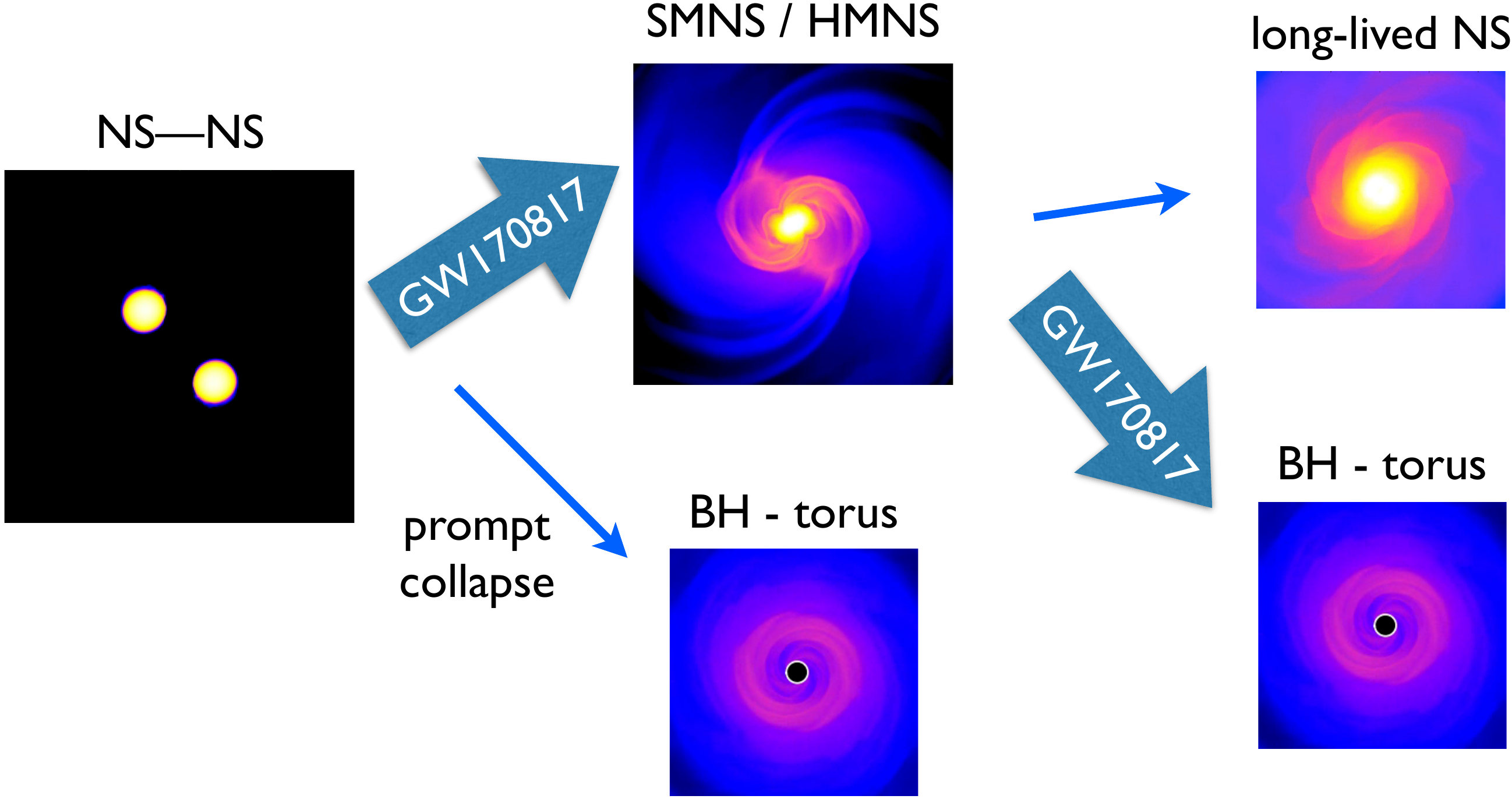}
\caption{Overview of the phenomenology for binary neutron star mergers, highlighting the most likely path for GW170817. The requirement for a metastable supramassive or hypermassive NS (SMNS/HMNS) of lifetime $\lesssim\!100\,\mathrm{ms}$ results from the combined requirement to maximize the amount of lanthanide-free ejecta by winds from such a remnant (in order to explain the blue kilonova emission of GW170817) and the need for lanthanide-rich outflows from a massive post-merger accretion disk (in order to power the red kilonova emission of GW170817; see the text for details; simulation snapshots courtesy of W.~Kastaun).}
\label{fig:GW170817_phenomenology}
\end{figure}

In conclusion, outflows from the remnant neutrino-cooled accretion disk provide a natural explanation for the combination of the high ejecta mass, low ejecta velocity, and low mean electron fraction of the red kilonova component of GW170817 (Eq.~\ref{eq:KN_properties}), and thus for the site of lanthanide production in GW170817. The requirement of a metastable remnant NS also aligns with the need to maximize the amount of shock-heated ejecta and with the need for winds from a remnant to explain the properties of the blue kilonova component (see above), which thus combine to a consistent picture. The resulting merger phenomenology is depicted in Fig.~\ref{fig:GW170817_phenomenology}. The detailed properties of winds from a remnant NS, however, remain largely uncertain, and the origin of the blue ejecta requires further investigation.

\subsection*{Could GW170817 have been a BH--NS merger?}

While the presence of electromagnetic counterparts in association with GW170817 necessitate matter and thus require at least one NS be present in the coalescing binary, the GW analysis alone could not rule out the possibility of a BH--NS merger \cite{LIGO+17DISCOVERY}. The measured component masses between $0.86-2.26\,M_\odot$ as well as the total mass of the binary of $2.73-3.29\,M_\odot$ \cite{LIGO+17DISCOVERY} are consistent with known galactic BNS systems, which have component masses $1.17-1.6\,M_\odot$ and total masses $2.57-2.88\,M_\odot$ \cite{Tauris2017}. Combined with the fact that the measured component masses are also significantly smaller than the BH masses found in galactic binary systems \cite{Ozel2010a,Kreidberg2012}, the GW observations of GW170817 favor a binary NS system. See Ref.~\cite{Yang2018}, however, for a discussion of processes that could form low-mass BHs. Regardless of astrophysical arguments about the existence of low-mass NS--BH systems, there is also an argument to be made based on the kilonova of GW170817, which also favors a binary NS system.

In BH--NS mergers, in which shock-heated ejecta and winds from a remnant are absent, the only known source of lanthanide-free (blue) ejecta are disk outflows from a post-merger accretion disk around the final BH as discussed above. As the dynamical BH-NS merger ejecta would likely be too small to explain the red kilonova of GW170817 \cite{Foucart2019}, the post-merger accretion disk would have to power both the lanthanide-free and lanthanide-bearing component of GW170817. Recent simulations of post-merger accretion disks in general-relativistic magnetohydrodynamics, which capture MHD turbulence to self-consistently describe angular momentum transport and thus plasma heating, indeed find a fraction of ejecta material with $Y_\mathrm{e}\gtrsim 0.25$ being ejected in the outflows of such accretion disks, mostly in polar directions \cite{Siegel2017a,Siegel2018a,Fernandez2019,Miller2019}. This is because in the non-degenerate, low-density and high-entropy polar regions, $e^+e^-$ pair creation is enhanced, which, in turn, increases positron capture $e^+ + n\rightarrow p + \bar{\nu}_e$, thus increasing the proton fraction $Y_\mathrm{e}$. However, we find substantial mixing in the outflows, such that an angular dependence and thus a two-component kilonova nature of the disk outflows is unlikely preserved to large spatial scales \cite{Siegel2018a} (but see \cite{Miller2019}). Furthermore, the amount of material with $Y_\mathrm{e}\gtrsim 0.25$ would be too small to explain the mass ratio between the blue and red kilonova components of GW170817 \cite{Siegel2017a,Siegel2018a,Fernandez2019}. Finally, the velocity of such blue disk winds $\sim\!0.1c$ would be inconsistent with the high velocities $v\approx 0.3c$ for the blue kilonova (Eq.~\eqref{eq:KN_properties}). Notwithstanding other lines of arguments based on BH masses, the inferred kilonova properties of GW170817 alone strongly disfavor a BH--NS merger origin.

\section{Implications and constraints on r-process nucleosynthesis}
\label{sec:implications}

\begin{figure*}[tb]
\centering
\includegraphics[width=0.8\textwidth]{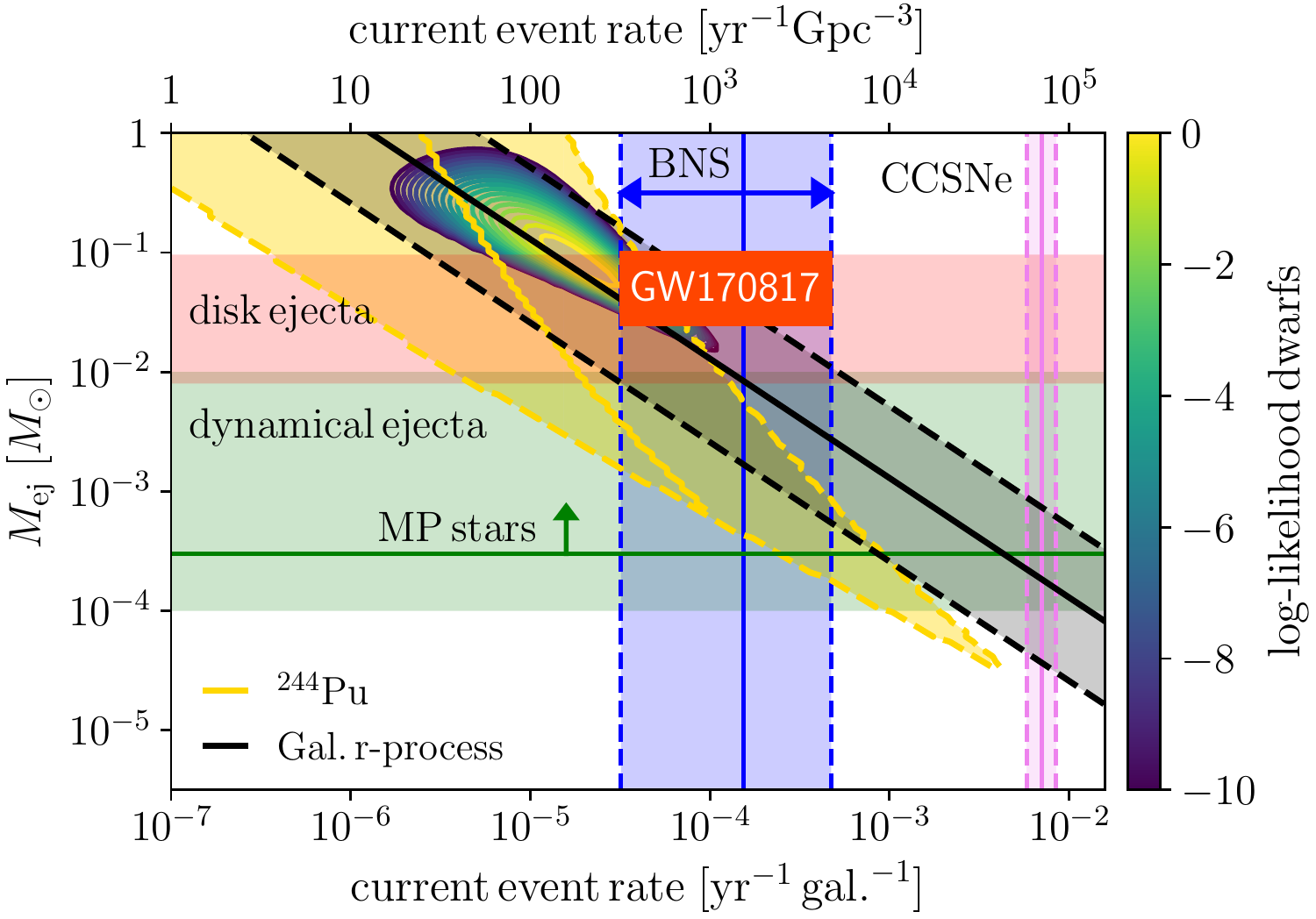}
\caption{Constraints on galactic r-process enrichment including neutron star mergers, in terms of ejecta mass per enrichment event versus galactic or current (local) volumetric event rate.  Shown are the local binary neutron star merger rate as inferred by LIGO/Virgo \cite{LIGO+17DISCOVERY}, and the local total core-collapse supernova rate for comparison \cite{Li2011}, with uncertainties indicated by dashed lines and colored bands. The ejecta mass for GW170817 represents the red (lanthanide-bearing) emission of the kilonova with uncertainties as discussed in Sec.~\ref{sec:EM_observations}. The horizontal red and green bands represent the range of ejecta masses for post-merger accretion disk outflows and dynamical ejecta in binary NS mergers as discussed in Sec.~\ref{sec:interpretation}. The yellow shaded region corresponds to constraints on r-process enrichment from measurements of $^{244}\mathrm{Pu}$ in the deep sea crust and from the early solar system \cite{Hotokezaka+15}. The color-coded log-likelihood contours depict rate-yield constraints from an analysis of dwarf galaxies in the halo of the Milky Way \cite{Beniamini2016a}, translated to r-process enrichment of the Milky Way disk. Also shown is the constraint from metal-poor halo stars of Ref.~\cite{Macias2018} (``MP stars''). The gray shaded region corresponds to the total amount of r-process material in the Milky Way disk. All mass estimates are computed assuming the solar r-process abundance pattern \cite{Arnould2007} starting at mass numbers $A\ge69$. See the text for details.}
\label{fig:BNS_rate_constraints}
\end{figure*}

In this section, we shall put the GW170817 kilonova observations and its interpretation into the broader context of galactic heavy r-process nucleosynthesis. Figure~\ref{fig:BNS_rate_constraints} compares the lanthanide-bearing (red) kilonova of GW170817 with masses and uncertainties as discussed in Sec.~\ref{sec:EM_observations} to various constraints on galactic heavy r-process nucleosynthesis. 

The comparison in Fig.~\ref{fig:BNS_rate_constraints} is formulated in terms of r-process mass ejected per enrichment event (being agnostic about the nature of the actual astrophysical mechanism or site) versus galactic or current/local ($z=0$) volumetric event rate. In computing the r-process ejecta masses, we assume a solar r-process abundance pattern \cite{Arnould2007} starting at mass number $A=69$. We note that for this choice of minimum atomic mass number and solar abundance pattern, the lanthanide mass fraction is consistent with that inferred for the red kilonova of GW170817. Volumetric rates are converted into galactic rates, assuming $\approx\!0.01$ Milky-Way equivalent galaxies per $\mathrm{Mpc}^{-3}$ \cite{Kopparapu2008}. Shown are the binary NS merger rate as measured by LIGO, $R_{\rm{BNS}}=1540^{+3200}_{-1220}\,\rm{Gpc}^{-3}\rm{yr}^{-1}$ \cite{LIGO+17DISCOVERY}, as well as the local total core-collapse supernova rate for comparison, $R_{\rm{CCSN}}=7.05^{+1.43}_{-1.25}\times 10^{4}\,\rm{Gpc}^{-3}\rm{yr}^{-1}$ \cite{Li2011}. Over-plotted as red and green horizontal bands are the range of ejecta masses obtained from numerical simulations for post-merger accretion disk outflows and dynamical ejecta, respectively (see Sec.~\ref{sec:interpretation}). 

\begin{figure}[tb]
\centering
\includegraphics[width=0.49\textwidth]{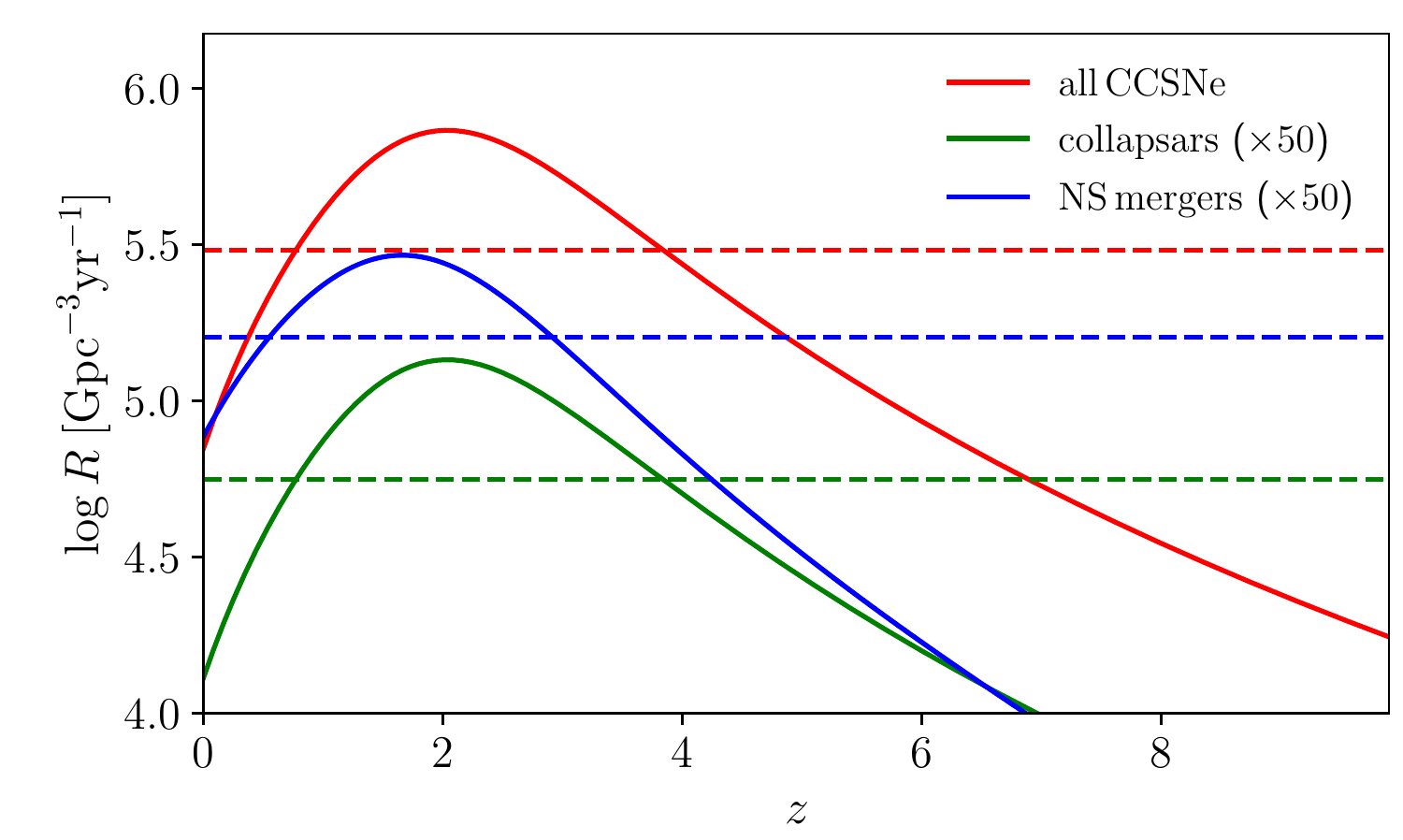}
\caption{Volumetric rates of core-collapse supernovae, neutron-star mergers, and collapsars as a function of redshift for the cosmic star formation history \cite{Madau2017}, assuming rates as defined in the model by \cite{Siegel2019a}. Average rates are indicated by dashed lines. Core-collapse supernovae (and collapsars, being a rare subclass thereof) follow the star formation history directly, while neutron-star mergers follow with a $\propto t^{-1}$ delay-time distribution.}
\label{fig:rates}
\end{figure}

The total amount of r-process material in the galaxy can be roughly estimated to $M_\mathrm{r,MW,A\ge69}\approx 2.6\times10^4\,M_\odot$, assuming a total stellar mass of the Milky Way of $(6.43\pm0.63)\times10^{10}\,M_\odot$ \cite{McMillan2011} and a mean r-process abundance pattern for stars in the Galactic disk similar to solar \cite{Arnould2007}. This assumes a `closed box' and thus, in particular, neglects the r-process material lost in Galactic outflows. Such outflows due to energy and momentum injection in star forming regions \cite{Heckman2000,Murray2005} can be sizable; in the case of the Galactic disk, the instantaneous mass loss rate is thought to be of the order of tens of percent of the star formation rate (e.g., \cite{Chisholm2017,Cote2016a}). Therefore, the requirement to produce the estimated current Galactic r-process content only represents a lower bound on Galactic r-process production, i.e., a necessary `test' any major r-process pollution source has to pass; the actual required amount of r-process material may be significantly higher.

The production of $M_\mathrm{r,MW}$ of r-process material in the Milky Way is, in principle, degenerate with respect to the average rate $R_\mathrm{MW}$ and yield per event $M_\mathrm{ej}$ \cite{Hotokezaka+15,Rosswog2017a,Hotokezaka2018a},
\begin{equation}
	M_\mathrm{r,MW} = M_\mathrm{ej} R_\mathrm{MW} t_\mathrm{MW}, \label{eq:M_r_MW}
\end{equation}
where $t_\mathrm{MW}\approx 10\,\mathrm{Gyr}$ is the age of the Milky Way. In translating this average rate over the lifetime of the Milky Way into a current rate one needs to make an assumption about the underlying enrichment process and the evolution of its rate as a function of redshift (i.e., cosmic time). For the purpose of this discussion, we shall assume that such enrichment can either be provided by NS mergers or by a rare subclass of CCSNe (such as collapsars \cite{Siegel2019a}, see Sec.~\ref{sec:collapsars}, or MHD supernovae, see Sec.~\ref{sec:MHD_supernovae}). We employ the calibrated rate model of Ref.~\cite{Siegel2019a} to obtain the corresponding evolution of rates as a function of redshift, assuming that CCSNe follow the star-formation history of the Milky Way with negligible delay, while NS mergers follow with a delay-time distribution $\propto t^{-1}$ and a minimum delay time of $20\,\mathrm{Myr}$; the results are depicted in Fig.~\ref{fig:rates}. The star-formation history of the Milky Way is assumed to follow the cosmic star formation history \cite{Madau2017}. We find that the current event rates for NS mergers and CCSNe (or collapsars) are, respectively, roughly a factor of two and five lower than their cosmic average. The degeneracy of producing the total r-process content in the Milky Way is shown in Fig.~\ref{fig:BNS_rate_constraints} by the black line and corresponding gray shaded region, assuming a factor of two uncertainty in the total ejecta mass estimate (due to order-unity uncertainties in the r-process pattern) and accounting for the fact that the current event rate may be lower with respect to the average event rate by a factor of up to five.

The rate-yield degeneracy in producing the total r-process content of the Milky Way is broken by measurements of the radioactive r-process isotope $^{244}\mathrm{Pu}$ with half-life of 81\,Myr \cite{Hotokezaka+15}. Suitable radioactive isotopes make it possible to break this degeneracy due to the fact that abundances of such isotopes in the inter-stellar medium (ISM) at a given location are determined by the local production history on the time scale of their mean lives (rather than by the entire galactic production history). The discrepancy in abundances of a factor of $\sim\!10^{-2}$ between measurements of $^{244}\mathrm{Pu}$ in the early solar system, based on abundance measurements of daughter nuclei in meteorites and ancient rocks \cite{Turner2007}, and the accumulation of $^{244}\mathrm{Pu}$ from the ISM onto the Earth's deep sea floor over the past $\approx\!25\,\mathrm{Myr}$ \cite{Wallner+15} requires low-rate, high-yield enrichment events, rather than high-rate, low-yield events \cite{Hotokezaka+15}. The allowed parameter space obtained in Ref.~\cite{Hotokezaka+15}, translated to an r-process pattern starting at $A=69$, is indicated as a yellow shaded region in Fig.~\ref{fig:BNS_rate_constraints}. This excludes ordinary CCSNe as potential contributors to the heavy r-process, corroborating previous theoretical arguments showing that the conditions for a strong (second-to-third peak) r-process are not met in such supernova environments \cite{Qian&Woosley96,Thompson+01,Roberts+12,MartinezPinedo+12}.

The rate-yield degeneracy is also broken by observations of strong r-process enrichment at low metallicity in dwarf galaxies in the halo of the Milky Way. The combination of strong upper limits on Eu production in some ultra-faint dwarf galaxies and strong enhancements of Eu in other ultra-faint dwarfs \cite{Ji2016,Hansen2017} as well as in other dwarf galaxies favors rare high-yield events ($\approx\! 1$ event per $10^{3}$ CCSNe) \cite{Beniamini2016a}. The likelihood analysis of \cite{Beniamini2016a} can be translated to a rate-yield constraint on r-process production in the Milky Way by normalizing to the current total CCSN rate in the Milky Way. Modeling the rates of CCSNe as discussed above, we obtain a mean CCSN rate for the Milky Way of $\approx\!3\times 10^{5}\,\mathrm{Gpc}^{-3}\,\mathrm{yr}^{-1}$. The translated log-likelihood contours are depicted in Fig.~\ref{fig:BNS_rate_constraints}.

In metal-poor halo stars of the Milky Way, whose abundances of $\alpha$-elements suggest they have been polluted by just a single CCSN, strong enhancements in Eu, again suggestive of only a single r-process pollution event, place a lower limit on the total r-process mass per enrichment event. Using the largest [Eu/H] enhancement among such stars and normalizing to the ISM mass swept up by the enrichment blast wave, Ref.~\cite{Macias2018} places a lower limit of $\approx3\times\!10^{-4}\,M_\odot$ per event \cite{Hotokezaka2018a} (translated to an explosion energy of $>10^{50}$ valid for both NS mergers and GRB supernovae, i.e., collapsars, and assuming an ISM density of $1\,\mathrm{cm}^{-3}$; indicated as ``MP stars'' in Fig.~\ref{fig:BNS_rate_constraints}). This constraint again favors high-yield events, and, in combination with the galactic yield (see Eq.~\eqref{eq:M_r_MW} above), low-rate events, disfavoring CCSNe. It depends on the explosion energy and on the density of the surrounding ISM, which is typically lower for NS mergers than for collapsars, as NS mergers tend to merge away from their birth sites due to systemic kicks received from the supernova explosions \cite{Berger2014,Beniamini2016a}.

Assuming that GW170817 was a typical binary NS merger in terms of ejecta masses, Fig.~\ref{fig:BNS_rate_constraints} shows that NS mergers are consistent with all constraints on r-process enrichment discussed above. In particular, Fig.~\ref{fig:BNS_rate_constraints} highlights how well ejecta from the post-merger accretion disk satisfy these constraints and provide a natural theoretical explanation for heavy r-process nucleosynthesis in such mergers. Therefore, in principle, NS mergers could account for most or even the entire amount of galactic heavy r-process material \cite{Kasen2017,Cote2018a,Hotokezaka2018a}. However, large uncertainties exist regarding both the NS merger rate and the exact quantity of ejecta (see Sec.~\ref{sec:EM_observations}). Additionally, as noted above, the constraint on the total Galactic r-process content neglects the effect of Galactic outflows and thus only represents a minimal requirement for total r-process nucleosynthesis; an additional source may be needed. Furthermore, at a closer look, there exist several challenges for NS merger models in explaining the Galactic r-process abundances, which we shall now discuss.

\section{Challenges for merger-only r-process enrichment}
\label{sec:challenges}

Galactic r-process enrichment solely based on NS mergers faces several challenges, which we shall outline below.

\subsection*{(i) Galactic r-process at low metallicities} 
It remains an open question whether NS mergers can account for the high Eu enrichment at low metallicities observed in stars of the Milky Way halo, as such mergers typically occur with a substantial delay with respect to star formation. While cosmological zoom-in simulations taking into account hydrodynamic mixing processes can explain the observed [Eu/Fe] abundance ratios and scatter at least at metallicities $[\mathrm{Fe/H}]\gtrsim -2$ \cite{vandeVoort2015,Shen2015}\footnote{For a discussion of the differences in Refs.~\cite{vandeVoort2015} and \cite{Shen2015} with regard to explaining the observed abundance ratios, see Refs.~\cite{Cowan2019,vandeVoort2019}.}, observed abundance ratios and scatter at lower metallicities ($[\mathrm{Fe/H}] < -2$) still remain difficult to explain \cite{vandeVoort2019}. Other studies taking inhomogeneous mixing into account cannot reproduce the observed amount and spread of [Eu/Fe] for $[\mathrm{Fe/H}]\lesssim -3$ to $[\mathrm{Fe/H}]\approx -2$ \cite{Cescutti2015,Wehmeyer2015}, unless NS binaries merge within an extremely short delay time of 10\,Myr \cite{Cescutti2015}. It was argued that the required r-process enhancement and scatter at $[\mathrm{Fe/H}]\lesssim -3$ can be obtained with more realistic NS merger delay times by assuming a hierarchical assembly of the Galactic halo by merging of sub-halos, if the star formation efficiencies are lower for less massive sub-halos \cite{Ishimaru2015,Hirai2015,Ojima2018} and if NS merger ejecta from one halo can cross-pollute other proto-galaxies and the intergalactic matter \cite{Komiya2016}; the higher metal loss rates due to shallower gravitational potentials of these sub-halos shift abundance features to lower metallicity.

\subsection*{(ii) r-process in ultra-faint dwarf galaxies} 
Recent observations of strong Eu-enhanced stars in ultra-faint dwarf galaxies in the halo of the Milky Way \cite{Ji2016,Hansen2017} challenge the NS merger scenario for r-process enrichment. This is because such dwarfs have very small escape speeds of $\sim\!10\,\mathrm{km}\,\mathrm{s}^{-1}$ and very short epochs of star formation of $\lesssim\!1\,\mathrm{Gyr}$. NS binaries must thus have both very small systemic kick velocities and short merger timescales of at most a few hundred Myr, in order to be retained in the galaxy and to feed r-process enriched material back into star formation. 

Ref.~\cite{Beniamini2016a} argues that a sizable fraction of such ``rapid mergers'' exist which have center of mass velocities comparable to the escape speed of the galaxy. The existence of such rapid mergers, however, strongly depends on the assumed initial binary separation (they are dominated by small-eccentricity systems whose timescale for merger is determined by the initial separation). Pre-supernova binary separations cannot be too tight, however, as surviving binaries would usually receive post-supernova systemic velocities that unbind such systems from their host galaxy. Furthermore, even with low kick velocities, the extremely shallow gravitational potentials lead to mergers at typically large offsets from star-forming regions, which results in dilution of r-process material available for internal enrichment, in tension with observed abundances \cite{Bonetti2019}. 

Another possibility for enrichment is through ultra-fast mergers from extremely hardened binaries that underwent a second phase of Case BB mass transfer \cite{Chruslinska2018} and merge before its post-supernova systemic velocity unbinds it from the galaxy \cite{Safarzadeh2019a,Zevin2019}.\footnote{Roche lobe overflow resulting from expansion of a helium-star after core helium depletion is referred to as Case BB mass transfer \cite{Savonije1976,DeGreve1977,Delgado1981,Tauris2013}. Such mass transfer is thought to give rise to extremely hardened binary neutron stars with orbital separations $\lesssim 0.1R_\odot$ and coalescense times $\gtrsim 1\,\text{kyr}$ \cite{Chruslinska2018,Vigna-Gomez2018}.} While the probability of such mergers might still be high enough to explain the Eu enhancement in some more massive ultra-faint dwarf galaxies such as Tucana III, the probability seems too low to explain enrichment in less massive ultra-faint dwarfs such as Reticulum II \cite{Safarzadeh2019a,Zevin2019}. In particular, Ref.~\cite{Zevin2019} finds that $\approx\!6\%$ and $\approx\!1\%$ of Tucana III-like and Reticulum II-like dwarfs would have an enrichment event, respectively, which is in tension with $\approx\!20\%$ of ultra-faint dwarfs showing r-process enhancements.

\subsection*{(iii) Globular clusters}

Globular clusters with large observed internal r-process abundance spreads \cite{Sneden1997,Roederer2011,Sobeck2011,Worley2013} are arguably even more challenging for NS merger enrichment models than ultra-faint dwarf galaxies discussed above. This is because of the very brief star formation periods ($\lesssim\!10-100\,\mathrm{Myr}$) of globular cluster progenitors, their shallow gravitational potentials with escape speeds of $10-100\,\mathrm{km\,s}^{-1}$ and very small physical sizes of only a few parsec (for recent reviews see, e.g., Refs.~\cite{Gratton2012,Bastian2018}). While dynamically formed binaries merge far too late and can thus be ruled out as r-process enrichment candidates \cite{Zevin2019}, ultra-fast mergers resulting from standard binary evolution through a Case BB mass transfer phase might be able to explain the observed fraction of r-process enriched globular clusters, albeit with some tension \cite{Zevin2019}. However, for such mergers to be viable enrichment candidates there must be dense, star-forming gas in the globular cluster $\sim\!30-50$ Myr after the initial burst of star formation \cite{Zevin2019}. This burst typically only lasts for $\approx\!10$ Myr, after which the first supernovae of massive stars efficiently remove the natal gas embedded in the globular cluster. Hence, scenarios for redistributing gas in such clusters on longer timescales need to be invoked, such as, e.g., winds from asymptotic giant branch stars \cite{Bekki2017}.

\subsection*{(iv) r-process in the Galactic disk}

\begin{figure}[tb]
\centering
\includegraphics[width=0.49\textwidth]{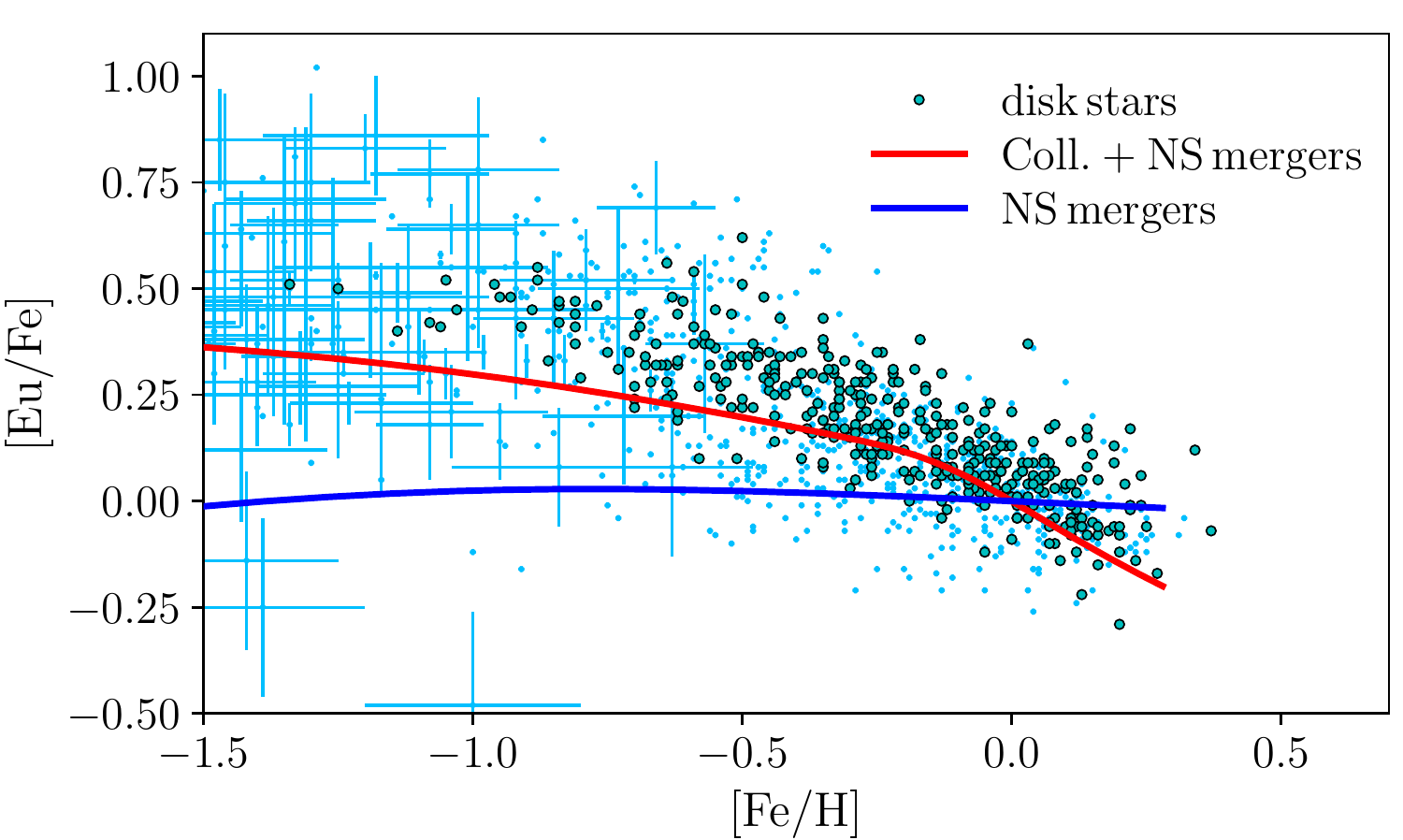}
\includegraphics[width=0.49\textwidth]{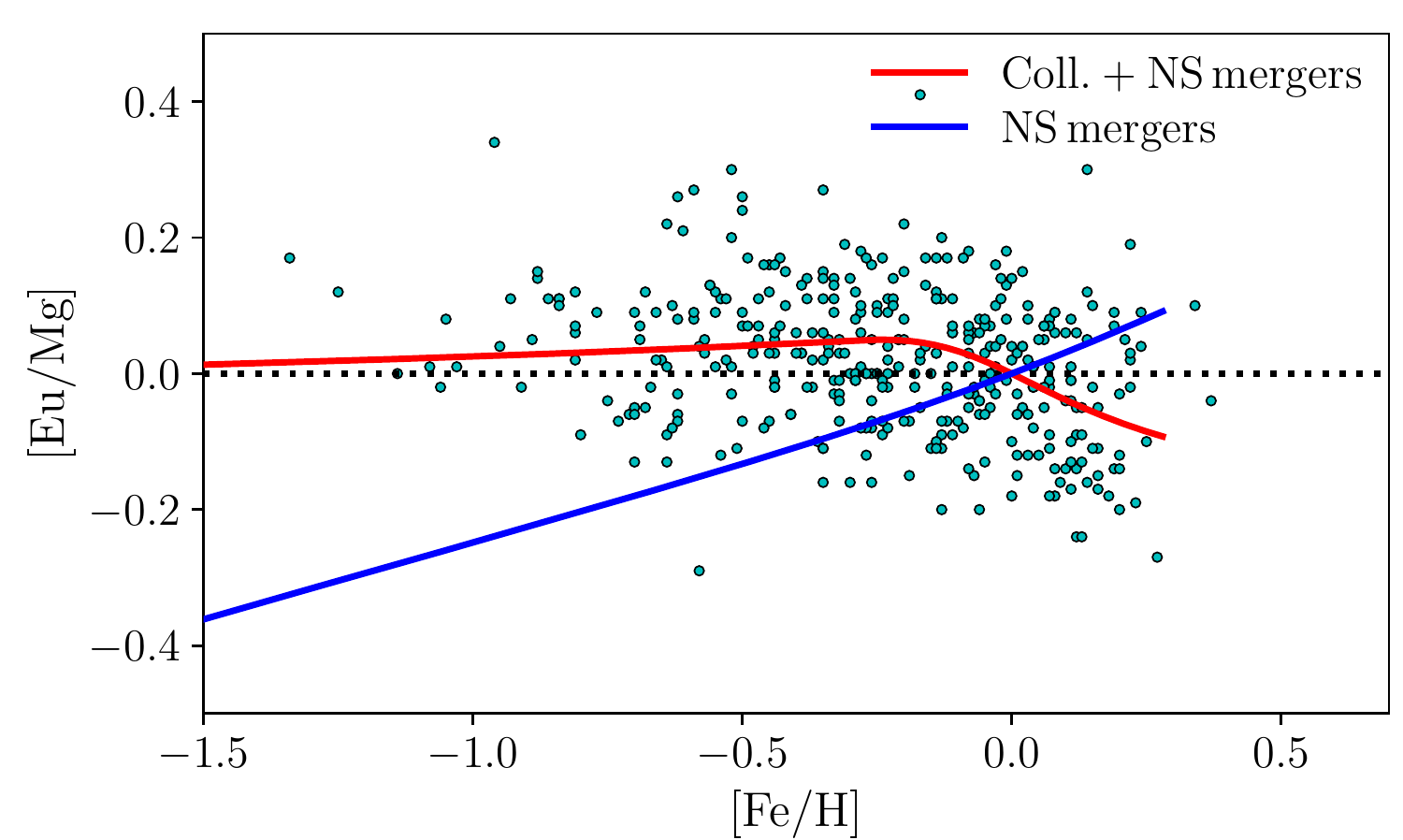}
\caption{Top: evolution of europium versus iron at high metallicity. Shown are observed abundances from the SAGA database \cite{SAGAdatabase,Suda2008} (light blue), and abundances specifically for Galactic disk stars \cite{Battistini2016} (blue dots). They are compared to predictions for the [Eu/Fe] evolution at late times using the one-zone model in Ref.~\cite{Siegel2019a}, assuming a minimum Supernova Type Ia delay time of 400 Myr. Predictions for merger-only r-process enrichment are shown in blue, while the fiducial combined model of NS mergers and collapsars of \cite{Siegel2019a} is shown in red, for which the dominant contribution comes from collapsars ($81\%$ at the time of formation of the solar system). The negative trend of [Eu/Fe] at late times cannot be reproduced with NS mergers alone, but instead requires enrichment events that more closely track the star formation rate, such as collapsars. Bottom: same as above, but showing the europium versus magnesium (alpha-element) trend for Galactic disk stars. Again, the trend is not reproduced with NS mergers alone, but can be obtained if the dominant contribution comes from collapsars.}
\label{fig:coll_NS_comp}
\end{figure}

It has been pointed out that the evolution of Eu relative to iron at high metallicities (in the Milky Way disk) appears to be inconsistent with r-process enrichment from NS mergers \cite{Cote2017b,Hotokezaka2018a} (see also \cite{Siegel2019a,Cote2018b}). At such late times, Fe is produced by Type Ia SN, which follow the same delay time distribution $\propto t^{-1}$ as NS mergers. As a result, the delay time distributions `cancel out' and the relative evolution of [Eu/Fe] is at best roughly flat or even rising. However, the observed trend of [Eu/Fe] with respect to metallicity [Fe/H] among disk stars has a negative slope. This is illustrated in Fig.~\ref{fig:coll_NS_comp}, which compares observed [Eu/Fe] abundances at high metallicity to predictions from a one-zone chemical evolution model \cite{Siegel2019a}. In order to obtain a decrease in [Eu/Fe], enrichment events that more closely follow the star formation history of the Milky Way are needed \cite{Siegel2019a,Cote2018b}, such as a rare subclass of CCSNe (e.g., collapsars; see Fig.~\ref{fig:coll_NS_comp} and below).

\subsection*{(v) r-process versus $\alpha$-elements in the Galactic disk}

Another challenge relates to the observed trend of r-process elements (such as Eu) relative to $\alpha$-elements (such as Mg) for stars in the Galactic disk. Ref.~\cite{Schoenrich2019} discusses the fact that a higher europium vs. Mg content relative to solar ($\mathrm{[Eu/Mg]}>0$) at metallicities of $[\mathrm{Fe/H}]\approx -1$ cannot be obtained in simple chemical evolution models for merger-only r-process enrichment. Essentially, r-process production always lags behind $\alpha$-element production by CCSNe due to the delay-time distribution, which precludes a negative trend at high metallicity. This is illustrated in Fig.~\ref{fig:coll_NS_comp} employing the one-zone chemical evolution model of Ref.~\cite{Siegel2019a}. In contrast, as shown in Fig.~\ref{fig:coll_NS_comp}, adding a dominant r-process contribution from collapsars (which track the star formation history, as $\alpha$-elements do), together with the fact that collapsars preferentially occur below a certain metallicity threshold that is slightly sub-solar (see the discussion in Ref.~\cite{Siegel2019a}), reproduces the observed [Eu/Mg] trend even in simple chemical evolution models, and for the same set of parameters as in (iv) (see also Sec.~\ref{sec:collapsars} and Fig.~\ref{fig:Mg_Eu_coll_NS}).

\section{GW170817 points to collapsars as the major source of r-process elements}
\label{sec:collapsars}

The realization that a NS post-merger accretion disk most likely ejected $\approx\!0.05\,M_\odot$ of heavy r-process material to generate the red kilonova emission of GW170817 (see Secs.~\ref{sec:EM_observations} and \ref{sec:interpretation}) suggests that similar physics may apply to the neutrino-cooled accretion disks in collapsars, i.e., the accretion disks that form around the final BH following the supernova-triggering collapse of rapidly-rotating massive stars, which are also thought to power long GRBs \cite{MacFadyen1999}. Indeed, recent general-relativistic magnetohydrodynamic simulations show that during the early stages of collapsar accretion (those that also power the GRB), disk wind ejecta are sufficiently neutron-rich and robustly synthesize both light and heavy r-process nuclei up to the third r-process abundance peak at mass number $A\sim 195$, in excellent agreement with the solar abundance pattern \cite{Siegel2019a}.

The contribution of collapsars to the galactic r-process can be estimated in several ways, both empirically and theoretically (see \cite{Siegel2019a} for details):
\begin{itemize}
	\item Knowing the lanthanide-rich r-process ejecta mass of GW170817 of $\approx\!0.05\,M_\odot$, one can estimate the contribution of collapsars relative to mergers purely empirically by using the isotropic-equivalent energies and rates of short GRBs (mergers) versus long GRBs (collapsars). Although collapsars occur less frequently than mergers, the longer duration of long GRBs compared to short GRBs and the correspondingly larger amount of accreted mass imply higher r-process ejecta masses for collapsars that overcompensate the lower event rate by a factor of 4--30. Calibration to GW170817 thus suggests that collapsars contribute by far more r-process material to the Galaxy than NS mergers (i.e., at least 80\%). This is also consistent with the requirement to reproduce the late-time [Eu/Fe] trend for stars in the Milky Way disk (see below and Fig.~\ref{fig:coll_NS_comp}). 

	\item An independent absolute estimate per event can be derived empirically assuming that collapsars are responsible for the entire Galactic r-process. Integrating over the Galactic star formation history and assuming that collapsars follow the rate of long GRBs\footnote{accounting for the fact that collapsars only occur up to a certain metallicity threshold \cite{Stanek+06,Perley+16}} one finds a per-event yield of $0.08-0.3\,M_\odot$ for typical parameters. Employing the GW170817 ejecta mass this estimate is consistent with the previous one.

	\item The per event contribution of collapsars can be theoretically estimated from the disk wind properties as probed by the simulations in combination with theoretical models for fallback accretion from the collapsing progenitor star, which typically yields $\mathrm{few}\times 10^{-2}\,M_\odot$ to $\sim\!1\,M_\odot$ of r-process material, consistent with the second and first estimate.
\end{itemize}
Thus, somewhat ironically, GW170817 points to collapsars being the dominant contributor to heavy r-process nucleosynthesis.

\begin{figure*}[tb]
\centering
\includegraphics[width=0.8\textwidth]{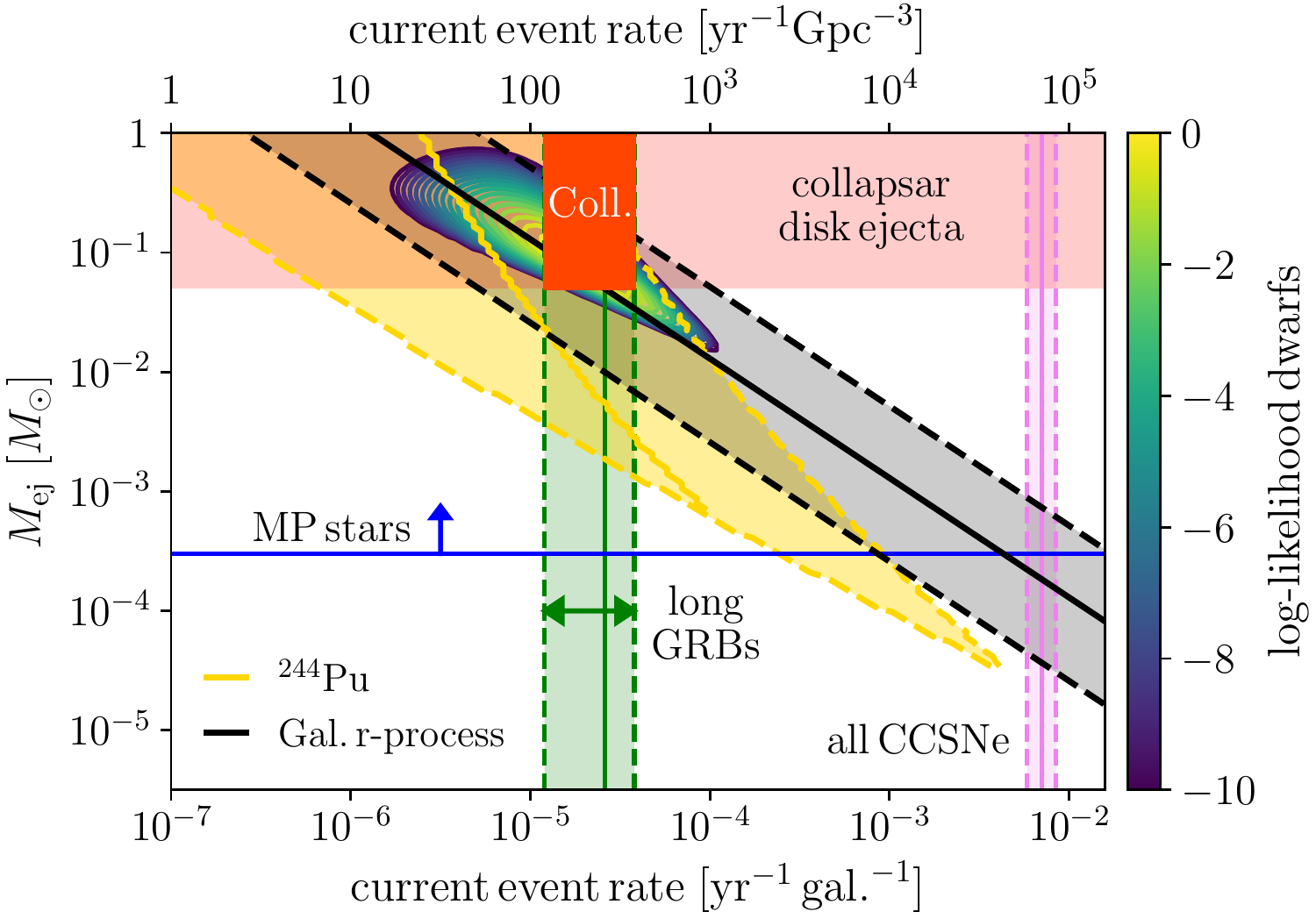}
\caption{Same as Fig.~\ref{fig:BNS_rate_constraints}, but including constraints from collapsars. The ejecta masses for heavy r-process material from collapsars are taken from \cite{Siegel2019a}, while the rate of collapsars (long GRB rate corrected for beaming) is taken from \cite{Wanderman&Piran10}, with a beaming fraction of $5\times10^{-3}$ \cite{Goldstein+16}.}
\label{fig:collapsar_rate_constraints}
\end{figure*}

Figure \ref{fig:collapsar_rate_constraints} presents a comparison of the collapsar scenario with the rate-yield constraints on heavy r-process nucleosynthesis, analogous to Fig.~\ref{fig:BNS_rate_constraints} for NS mergers. We note that although GRBs currently do not occur anymore in our galaxy due to a metallicity threshold \cite{Stanek+06,Perley+16,Palmerio2019}, rates are expressed in terms of an effective current rate using the rate model discussed above (see Fig.~\ref{fig:rates}) for comparison with the NS merger case (Fig.~\ref{fig:BNS_rate_constraints}). As evident from Figs.~\ref{fig:BNS_rate_constraints} and \ref{fig:collapsar_rate_constraints}, the somewhat lower rate and higher yield per event of the collapsar scenario compared to NS mergers is in arguably better agreement with the constraints on r-process enrichment. As noted in Sec.~\ref{sec:implications}, consistency with these constraints implies that, in principle, based on the average rate and yield, collapsars qualify as a candidate for the dominant Galactic r-process site. Given the large uncertainties in both rates and yields for NS mergers and collapsars, and due to the fact that Figs.~\ref{fig:BNS_rate_constraints} and \ref{fig:collapsar_rate_constraints} neglect r-process material lost in Galactic outflows, this does not mean that only one of the two processes can be operating. Rather, both enrichment mechanisms may be contributing significantly to the Galactic r-process. At a closer look, however, important differences between the two r-process sites exist when comparing collapsars to the challenges outlined in Sec.~\ref{sec:challenges}.

\begin{figure}[tb]
\centering
\includegraphics[width=0.49\textwidth]{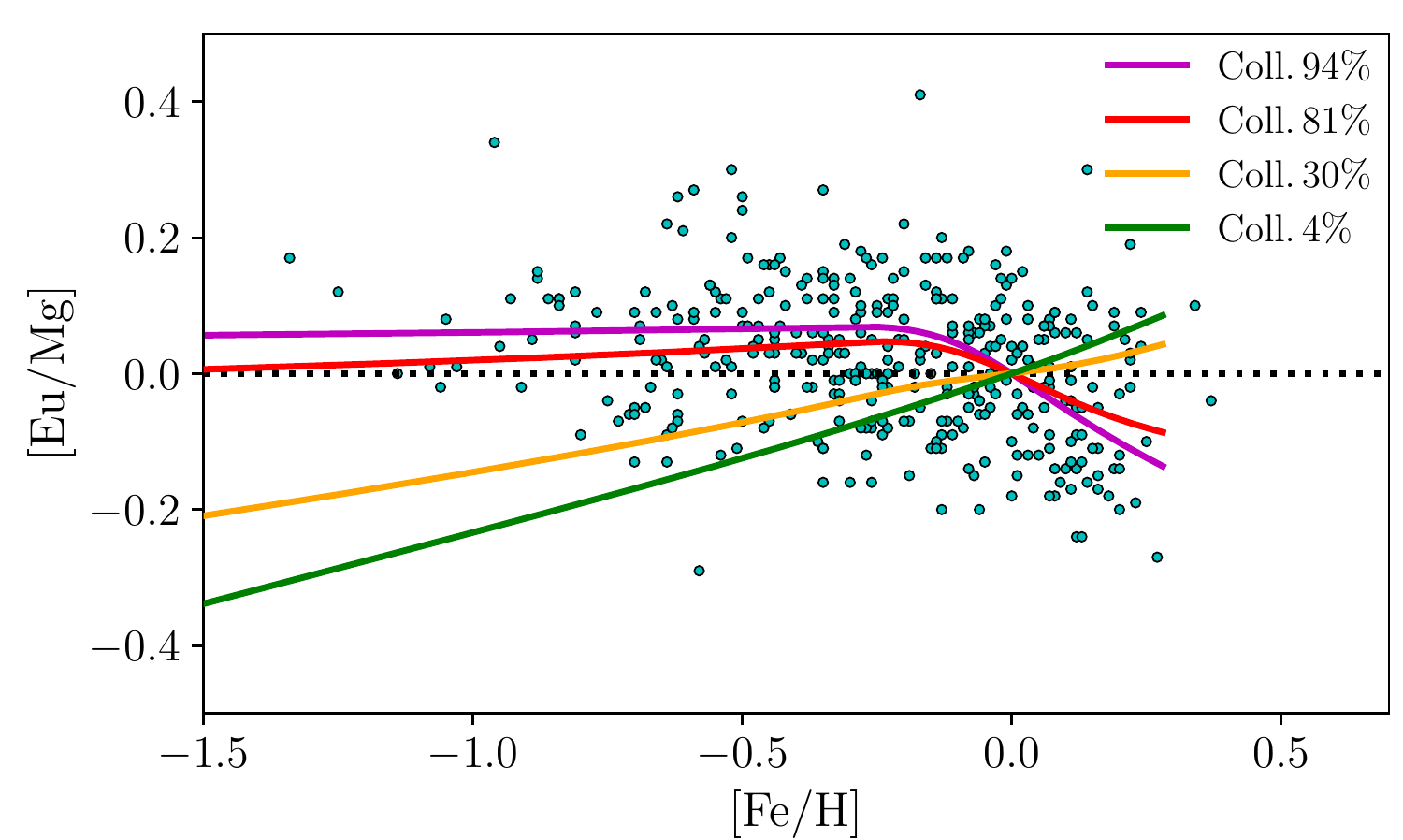}
\caption{Evolution of heavy r-process versus $\alpha$-elements at high metallicity. Shown are observed europium vs. magnesium abundances for galactic disk stars \cite{Battistini2016} (blue dots). They are compared to predictions for the [Eu/Mg] evolution at late times using the one-zone model in Ref.~\cite{Siegel2019a}, assuming a minimum Supernova Type Ia delay time of 400 Myr. Predictions for models with combined NS merger and collapsar contributions to r-process enrichment are shown, labelled by the contribution of collapsars to the total at the time of formation of the Solar System. The calibrated fiducial model of Ref.~\cite{Siegel2019a} (also shown in Fig.~\ref{fig:coll_NS_comp}) corresponds to the red curve. The observed mean trend can only be obtained with a dominant contribution from collapsars.}
\label{fig:Mg_Eu_coll_NS}
\end{figure}

Collapsars either do not face or appear to resolve the observational challenges for NS merger models discussed in Sec.~\ref{sec:challenges}. This is because they result from the core-collapse of massive stars with only a few Myr lifetime and occur directly in the star-forming region that gave birth to the massive star. This is consistent with the observation that long GRBs follow the radial distribution expected for star formation in disk galaxies \cite{Bloom2002} and that they are spatially correlated with bright star-forming regions within their host galaxies \cite{Fruchter2006}. As a result, r-process enrichment through collapsars does not suffer from uncertainties and problems associated with binary stellar evolution, binary kick velocities, and merger delay times in various environments---challenges (i)-(iii) listed in Sec.~\ref{sec:challenges} are bypassed. Collapsars as a rare type of core-collapse supernovae provide a source of prompt enrichment for halo stars that is in better agreement with current observations than neutron star mergers \cite{vandeVoort2019} (but see also Ref.~\cite{Macias2019} for constraints in closed `single-supernovae' environments). Prompt enrichment without spatial dislocation is also favorable for r-process enrichment in ultra-faint dwarfs and globular clusters \cite{Bonetti2019,Zevin2019}. In order to illustrate how collapsars appear to resolve challenges (iv) and (v), we again employ the one-zone chemical evolution model from Ref.~\cite{Siegel2019a}, allowing for contributions from both NS mergers and collapsars. As collapsars follow the star formation history without significant delay, a dominant contribution by collapsars to the galactic r-process reproduces the late-time decrease of europium versus iron in the Milky Way disk (cf.~Fig.~\ref{fig:coll_NS_comp}). Additionally, for the same set of parameters, the mean trend of r-process versus $\alpha$-elements for Galactic disk stars is obtained (cf.~Fig.~\ref{fig:coll_NS_comp}). As a final remark, collapsars occur mostly in low-metallicity environments \cite{Stanek+06} and would thus be over-represented among the first generations of stars, which provides a natural explanation for the observed carbon-enhanced metal-poor (CEMP) stars with high r-process enrichment (e.g., \cite{Sneden2003}).

Observationally, one may hope to see direct evidence of r-process nucleosynthesis from collapsars in the late-time NIR lightcurves and spectra of nearby long GRB (broad-lined Type Ic) supernovae. Such observational signatures exist at late times, depending on how efficiently the r-process material from the collapsar disk winds mix into the supernova ejecta \cite{Siegel2019a}. Detection of such features will benefit from future observations of NS mergers with sensitive NIR telescopes, such as the James Webb Space Telescope, which will help to identify unique signatures (line features) of r-process nuclei in nebular spectra of kilonovae (`pure r-process sources'). Such observations can then be compared to late-time GRB supernovae (`giant kilonovae in supernovae') to help identify the presence of r-process material.

\section{MHD supernovae}
\label{sec:MHD_supernovae}

Another contender for heavy r-process enrichment are MHD supernovae \cite{Winteler2012,Thompson+04,Metzger+08}. While some simulations found that magnetized jets in such supernova explosions can give rise to a sizable amount of fast-expanding neutron-rich ejecta \cite{Winteler2012,Nishimura2015,Nishimura2017}, these events are challenged to eject significant amounts of heavy r-process nuclei (mass numbers $A\gtrsim 130$) when considering the three-dimensional stability of the jets \cite{Moesta2014b,Moesta2018,Halevi&Mosta18}. Additionally, if MHD supernovae did produce significant amounts of heavy r-process elements, the high opacity of the lanthanide material would be mixed with the $^{56}\mathrm{Ni}$ of the supernova ejecta in a way that would likely be incompatible with present observations (see Fig.~3 in \cite{Siegel2019a}). 

\begin{figure}[tb]
\centering
\includegraphics[width=0.49\textwidth]{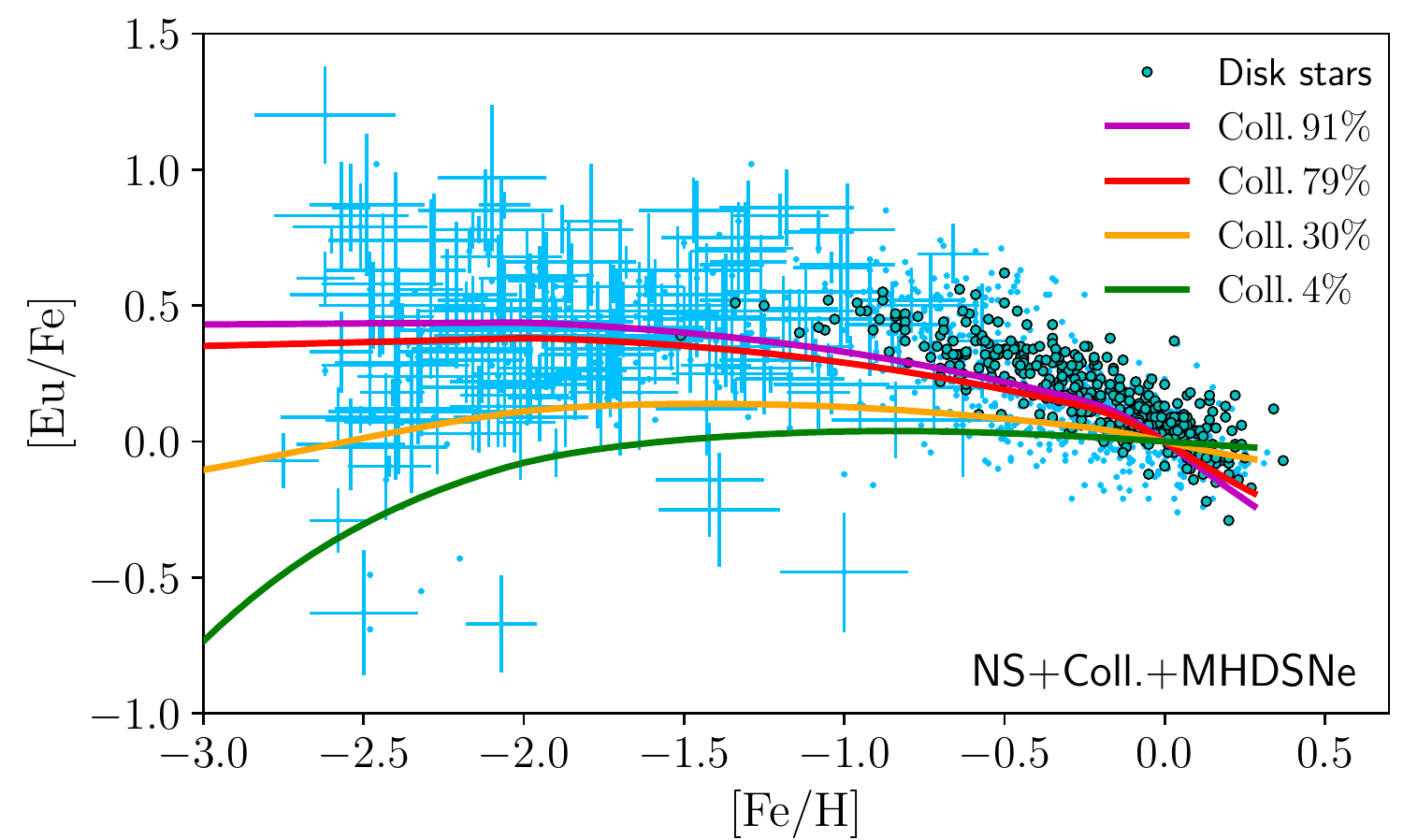}
\includegraphics[width=0.49\textwidth]{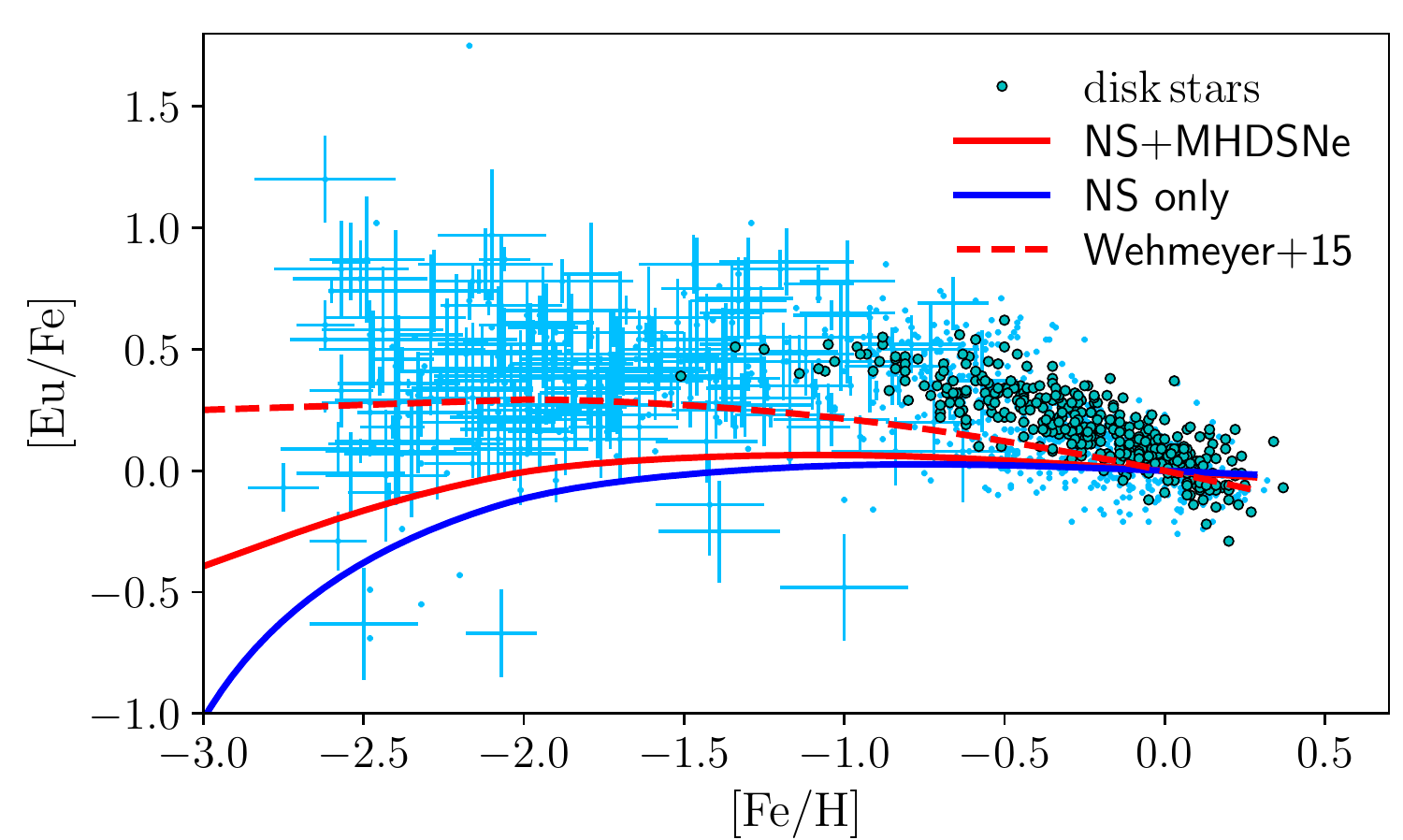}
\caption{Top: evolution of heavy r-process elements versus iron as a function of metallicity for scenarios including NS mergers, collapsars, and optimistic MHD supernova yields and rates. Observed abundances are from the SAGA database \cite{SAGAdatabase,Suda2008} (light blue) and from \cite{Battistini2016} specifically for Galactic disk stars blue dots). The fiducial NS merger enrichment rate (intrinsic merger rate times enrichment efficiency) is varied between a factor of 0.3 and 100, and the resulting model curves are labeled according to the overall contribution of collapsars to the total Galactic r-process content at the time of formation of the Solar System (see Tab.~\ref{tab:enrichment_fractions} for detailed contributions of all enrichment sources). The contribution of MHD supernovae is always sub-dominant with respect to collapsars (see the text for details). The fiducial model (calibrated NS merger and collapsar rates as in Ref.~\cite{Siegel2019a}) correspond to the red curve. Bottom: calibrated fiducial model run as above, but setting the collapsar contribution to zero (red) or only retaining the NS merger contribution (blue). This is compared to a run including NS merger and MHD supernovae with rates and yields from Ref.~\cite{Wehmeyer2015} (red-dashed curve).}
\label{fig:Eu_Fe_coll_NS_MHDSN}
\end{figure}

Notwithstanding these issues, we shall here explore the consequences for Galactic r-process enrichment \emph{if} MHD supernovae do indeed produce significant amounts of lanthanides. To this end, we modify the Galactic chemical evolution model of Ref.~\cite{Siegel2019a} and add MHD supernovae as an additional enrichment source. We assume that MHD supernovae occur at $0.3\%$ of the total CCSNe rate, the higher end of the rate range considered by Ref.~\cite{Wehmeyer2015}, and assign a Eu yield of $1.4\times 10^{-5}\,M_\odot$ per event \cite{Winteler2012,Wehmeyer2015}. We assume that collapsars contribute $0.3\,M_\odot$ of iron and $0.3\,M_\odot$ of r-process material per event \cite{Siegel2019a}. As in Ref.~\cite{Siegel2019a} and Fig.~\ref{fig:Mg_Eu_coll_NS}, we vary the neutron-star merger contribution by renormalizing the fiducial neutron-star merger rate calibrated to the LIGO-Virgo observations (product of intrinsic rate and enrichment efficiency) by a factor between 0.3 and 100. Figure \ref{fig:Eu_Fe_coll_NS_MHDSN} (top) shows the resulting combined Galactic evolution of r-process enrichment from collapsars, NS mergers, and MHD supernovae; Tab.~\ref{tab:enrichment_fractions} lists the corresponding enrichment fractions of each enrichment scenario to the total r-process content of the Galaxy at the time of formation of the Solar System. Similar to the NS merger contribution, there are significant uncertainties in the collapsar rate and yield. However, even when using pessimistic collapsar rates and yields together with optimistic MHD SNe rates and yields, we generally find that contributions from MHD supernovae are sub-dominant with respect to collapsars.

The bottom panel of Fig.~\ref{fig:Eu_Fe_coll_NS_MHDSN} presents a comparison of the above model to previous work (e.g., \cite{Wehmeyer2015}) that concluded that MHD supernovae can provide important contributions to Galactic europium at low metallicities. We neglect any contributions from collapsars and plot model runs retaining NS mergers and MHD supernovae only as well as a model run with NS mergers only for reference (red and blue curves, see bottom panel of Fig.~\ref{fig:Eu_Fe_coll_NS_MHDSN}). Those are compared to a run using the assumed NS merger rate and yields from Ref.~\cite{Wehmeyer2015} ($1.28\times 10^{-2}\,M_\odot$ of total ejecta per merger, including $10^{-4}\,M_\odot$ europium, occuring at a present probability of $3.4\times10^{-4}$ of the total CCSNe rate or $24\,\mathrm{Gpc}^{-3}\,\mathrm{yr}^{-1}$), see red-dashed curve in the bottom panel of Fig.~\ref{fig:Eu_Fe_coll_NS_MHDSN}. While the contribution of MHD supernovae with respect to NS mergers only is small in our model, it is much more pronounced when adopting the NS merger parameters of Ref.~\cite{Wehmeyer2015}. This is due to decreasing the NS merger contribution relative to MHD supernovae when adopting the latter parameters. Again, it should be emphasized that the simple chemical evolution model employed here assumes instantaneous mixing and therefore cannot predict scatter in abundances, it can only roughly predict the expected mean abundance trend. In short, the conclusion drawn here that if MHD supernovae produce significant europium their contribution would likely be minor (even if collapsars are neglected) is not in contradiction to previous work; it rather follows from updated NS merger rates and yields, at least to leading order.

\begin{table}
\caption{Fraction of overall r-process material contributed to the Galaxy by NS mergers (NSMs), collapsars, and MHD supernovae (MHD SNe) at the time of formation of the Solar System. In order to illustrate different scenarios, the NSM contribution is varied by changing the product of intrinsic rate and enrichment efficiency $f_\mathrm{NS}R_\mathrm{NSNS}$, where the fiducial intrinsic rate is $1540\,\mathrm{Gpc}^{-3}\mathrm{yr}^{-1}$ \cite{LIGO+17DISCOVERY}. The cases listed here are plotted in Fig.~\ref{fig:Eu_Fe_coll_NS_MHDSN} (see the text for details).}
\label{tab:enrichment_fractions}
\begin{center}
\begin{tabular}{cccc}
\hline
$f_\mathrm{NS}R_\mathrm{NSNS}$ & NSMs & Collapsars & MHD SNe \\
$[0.5\times 1540\,\mathrm{Gpc}^{-3}\mathrm{yr}^{-1}]$ & [\%] & [\%] & [\%] \\
\hline
0.3 & 6.1 & 91.0 & 2.9\\
1.0 & 18.7 & 78.7 & 2.5 \\
10 & 69.4 & 29.6 &  1.0\\
100 & 95.7 & 4.2 &  0.1\\
\hline
\end{tabular}
\end{center}
\end{table}

\section{Light r-process elements and universality}
\label{sec:light_r_process}

While the discussion has so far mainly focused on the origin of lanthanide-bearing material, we shall now also briefly comment on r-process elements of the first-peak and in-between first and second peak (light or `limited' r-process elements, with atomic numbers $Z<50$ or mass numbers $A<115$). In the following, I briefly point out how both sites for heavy r-process element production discussed here (NS mergers and collapsars) can give rise to intrinsic scatter in the light r-process elements.

It is interesting to note that the abundances of such light r-process elements in stars of the ultra-faint dwarf galaxy Reticulum II and some extremely metal-poor stars (e.g., CS 22892-052) are significantly depleted with respect to the solar system r-process abundances \cite{Ji2016,Sneden2000,Sneden2003}. In general, while abundances of heavy r-process elements (lanthanides; $Z>56$) in metal-poor stars in the Milky Way halo and ultra-faint dwarfs show very little deviation from the scaled solar pattern (a `robust, quasi-universal' pattern from the second to the third r-process peak), there is significant scatter among the abundances of light r-process elements when scaled to heavier elements such as Eu, with some metal-poor stars showing higher first-peak abundances than solar (up to 1.5 dex scatter around the solar pattern; see, e.g., the compilation of abundance patterns in Ref.~\cite{Frebel2018}). The origin of this scatter remains unclear, but has been interpreted as being suggestive of lighter and heavier r-process elements being produced at different r-process sites by different types of r-processes (with the light elements being produced by a ``weak r-process'' or ``light-element primary process'') \cite{Sneden2000,Travaglio2004,Kratz2007,Siqueira2014}.

In NS mergers, the production of exclusively light r-process elements ($Z<56$) manifests itself in early blue kilonova emission, such as, most likely, the blue kilonova emission in GW170817 (see Sec.~\ref{sec:EM_observations}), as a non-negligible lanthanide fraction would strongly increase the opacity of the material and redden the transient \cite{Kasen2013,Barnes2013,Kasen2017}. The amount of such `blue' material depends on the amount of shock-heated ejecta, which can vary by orders of magnitude depending on the mass ratio of the binary and the EOS (see the discussion in Sec.~\ref{sec:interpretation}). It also depends on the amount of mass lost through winds form a remnant NS, which, depending on the lifetime of the object, can again vary by orders of magnitude (cf.~Sec.~\ref{sec:interpretation}). Finally, a long-lived remnant ($\gtrsim\!100\,\mathrm{ms}$) can also turn a significant fraction of the post-merger accretion disk winds into blue ejecta via its strong neutrino radiation \cite{Metzger2014c,Lippuner2017a,Fahlman2018}. These ejecta components superimpose with light r-process elements produced in the post-merger disk outflows, which possibly somewhat under-produce the first-peak elements relative to the solar pattern \cite{Siegel2018a,Miller2019}. Finally, also the amount of lanthanide-rich ejecta (disk and tidal ejecta) and thus the relative amount of light versus heavy r-process elements depends on binary parameters (binary mass ratio, total mass) and the EOS (see Sec.~\ref{sec:interpretation}).

As pointed out in Ref.~\cite{Siegel2019a}, similar to NS post-merger accretion disks collapsar accretion disks may give rise to an r-process abundance pattern in good agreement with the solar pattern (although possibly slightly under-pro\-duc\-ing first-peak elements). Variations regarding light r-process elements on top of this pattern result from a period of collapsar fallback accretion at intermediate accretion rates that additionally synthesizes $\sim\!\mathrm{few}\times 10^{-2}-\mathrm{few}\times 10^{-1}\,M_\odot$ of exclusively light r-process material (due to a reduced free neutron-to-seed ratio; cf.~Table 1 of \cite{Siegel2019a}). Note also that the ratio of this light r-process contribution to that of the main r-process material strongly varies, depending on the exact stellar structure of the progenitor (cf.~Table 1 of \cite{Siegel2019a}). Furthermore, during the early GRB SN, launched from a rapidly-spinning and strongly-magnetized proto-neutron star prior to collapse to a BH \cite{Winteler+12,Thompson+04,Metzger+08}, recent simulations accounting for the three-dimensional stability of magnetized jets find that a moderate quantity of $\lesssim 10^{-2}\,M_\odot$ of light r-process material could additionally be ejected \cite{Moesta2018,Halevi&Mosta18}. For particularly strong magnetic fields, a fraction of this material could also form lanthanides, but would be sub-dominant with respect to the collapsar disk ejecta. Significant amounts of heavy r-process elements from the GRB SN are likely ruled out based on GRB lightcurves and spectra \cite{Siegel2019a}. In conclusion, different fallback accretion histories resulting from different progenitor stars as well as diversity in the GRB supernova explosion, depending on progenitor properties (such as magnetic fields), naturally give rise to variations in the light r-process element abundances.

As discussed above, both collapsars and NS mergers can account for variations in the light r-process elements by orders of magnitude. Therefore, explaining the observed abundance variations in metal-poor stars does not necessarily require the light and heavy r-process elements to be produced in different astrophysical phenomena. Rather, these variations can naturally be produced in the same astrophysical phenomenon by intrinsic variations in the ejecta properties owing to a complex phenomenology of the events, which will be interesting to test with future (kilonova) observations \cite{Ji2019}. Universality among the heavy r-process elements, instead, may be obtained from a combination of fission \cite{Korobkin2012,Rosswog+14} and the fact that the nucleosynthesis conditions for heavy r-process elements in disk outflows are expected to be very similar both among individual collapsar and NS merger events and when comparing NS merger to collapsar events\footnote{If the EOS constraints deduced from GW170817 indeed hold true, most binary NS mergers are expected to form massive post-merger accretion disks \cite{Radice2018a,Radice2018b}; outflows from such disks would then dominate the merger ejecta when averaged over the binary NS merger population.}. The latter aspect is due to a self-regulation mechanism in such accretion disks that tends to regulate the disk into a state of mild electron degeneracy \cite{Siegel2017a,Siegel2018a}; this results in very similar heavy r-process nucleosynthesis patterns across a wide range of accretion rates above a critical accretion rate for r-process production (cf., e.g., Fig. 1 in \cite{Siegel2019a}).

\section{Conclusion and outlook}
\label{sec:conclusion}

GW170817 marked the beginning of a new era in astronomy, astrophysics, and nuclear physics. The spectacular kilonova associated with GW170817, whose properties we have discussed here, represents the first direct observation of cosmic r-process nucleosynthesis. Its photometric and spectral evolution is best described by a multi-component (`blue-red') kilonova, which is also expected on theoretical grounds for reasons discussed in detail in Sec.~\ref{sec:interpretation}. 

The observationally inferred kilonova parameters, accounting for realistic uncertainties (Sec.~\ref{sec:EM_observations}), suggest post-merger accretion disk outflows as the astrophysical production site of the heavy r-process elements (lanthanide-bearing ejecta) in this event \cite{Siegel2017a,Siegel2018a,Fernandez2019}. As discussed, such outflows provide a natural explanation for the combination of high ejecta mass, small ejecta velocity, and high neutron richness of the observed red kilonova emission, which none of the other known types of merger ejecta simultaneously satisfy. The blue kilonova emission most likely originated in a combination of shock-heated ejecta and winds from a temporarily stable remnant NS. The observed blue emission after the first data point 11\,h post-merger does require radioactive heating from a significant amount of blue merger ejecta (Eq.~\ref{eq:KN_properties}), as energy input at the time of merger due to cocoon heating \cite{Piro2018} would have been degraded due to adiabatic expansion \cite{Metzger2018a}.

The interpretation of the blue-red kilonova generated by shock-heated ejecta and winds from a remnant NS (blue emission), as well as outflows from a post-merger accretion disk (red emission), requires a soft EOS (small NS radii). As discussed, this simultaneously maximizes the amount of shock-heated ejecta, leads to a temporarily stable remnant NS (cf.~Fig.~\ref{fig:GW170817_phenomenology}) that launches additional winds, and leaves a massive post-merger accretion disk. The lifetime of the remnant NS is constrained to be at least a few tens of ms, in order to allow for significant mass ejection through winds to occur, and less than hundreds of ms, as otherwise strong neutrino radiation from the remnant would turn a significant fraction of the disk outflows into lanthanide-free (blue) ejecta. The upper bound on the lifetime is also consistent with the inferred kinetic energy of the ejecta, which would have been significantly higher, as the remnant inevitably transfers energy to the ejecta on longer timescales \cite{Margalit&Metzger17}. This consistent picture of the red and blue kilonova also has important implications for the EOS, as it puts tight constraints on the maximum mass of neutron stars (e.g., \cite{Margalit&Metzger17}) as well as on the tidal deformability \cite{Radice2018a}.

While, in principle, NS mergers could account for most or all of the galactic r-process elements (assuming that GW170817 was a typical merger), we have discussed how GW170817 actually points to collapsars as the main contributor to the galactic r-process \cite{Siegel2019a} (Sec.~\ref{sec:collapsars}). Compared to NS mergers, collapsars arguably better satisfy existing constraints on heavy r-process nucleosynthesis, such as constraints from $^{244}\mathrm{Pu}$ \cite{Hotokezaka+15}, Eu enrichment in dwarf galaxies \cite{Beniamini2016a}, the rate-yield constraints from the total galactic r-process content, and the respective event rate constraints (cf.~Secs.~\ref{sec:implications}, \ref{sec:collapsars} and Figs.~\ref{fig:BNS_rate_constraints}, \ref{fig:collapsar_rate_constraints}). The collapsar scenario also bypasses and solves a number of problems that afflict merger-only enrichment models (Sec.~\ref{sec:challenges}) for the Galactic r-process in various environments, both at low and high metallicities.

There are a few caveats regarding the interpretation discussed here. Neutron star mergers are so far the only enrichment source with direct observational evidence for r-process nucleosynthesis; direct observational evidence for heavy element production in collapsars and other contenders such as MHD SNe is still to be gathered. Further work is required to assess the robustness of the nucleosynthesis products of all enrichment sources (NS mergers, collapsars, MHD SNe) with respect to weak interactions and neutrino transport. Effects of magnetic fields also introduce uncertainty regarding the nucleosynthesis products. While this is less of an issue for neutron star post-merger accretion disks and collapsars, where magnetic fields are self-consistently amplified by the magnetorotational instability from arbitrarily weak magnetic seed fields, the assumed magnetic field strengths in the progenitor core appears to be crucial for predicting r-process abundances in MHD SNe; stable jets that lead to lanthanides production likely require unrealistically high progenitor magnetic field strengths with the current treatment of neutrino interactions \cite{Moesta2018}. Finally, more detailed chemical evolution studies are required to reflect on the consequences and importance of different enrichment mechanisms in the future; however, the simple model employed here is well suited to illustrate the issues and open questions.

GW170817 marked the beginning of a bright future for the study of cosmic nucleosynthesis. If the EOS is indeed rather soft as suggested by GW170817, then massive post-merger accretion disks are a ubiquitous phenomenon \cite{Radice2018a,Radice2018b}, and one would expect many future mergers to show massive and slow red kilonova components \cite{Siegel2018a} (assuming that the NS masses are consistent with the galactic distribution of double NS systems\footnote{Note that whether or not the merger leads to a prompt collapse to a BH (and thus to the absence of a massive post-merger accretion disk) depends on the total binary mass relative to the maximum NS mass.} \cite{Kiziltan2013}). In this sense, GW170817 would be a typical merger. However, GW170817 is only one specific case, and its interpretation is based on emission seen from one particular viewing angle. We will soon observe qualitatively different scenarios from various viewing angles and start to understand the rich phenomenology of binary NS mergers and, hopefully, of BH--NS mergers and their implications for r-process nucleosynthesis. We expect to see qualitatively new features in future events, examples of which are very early blue emission due to a neutron precursor \cite{Metzger+15} or energy injection from a relativistic jet into merger ejecta (cocoon heating) \cite{Gottlieb+18,Kasliwal2017}. Significantly unequal mass mergers could lead to strong tidal ejecta components, which if sufficiently neutron-rich for fission cycling may lead to actinide over-production \cite{Holmbeck2019}. This may provide insights into the origin of actinide-boost stars, which comprise roughly 30\% of all metal-poor stars with strong r-process enhancement \cite{Schatz2002,Roederer2009,Mashonkina2014}. Furthermore, the diversity of merger events may also help us understand scatter in the abundances of light r-process elements (first-to-second peak elements; cf.~Sec.~\ref{sec:light_r_process}), as well as provide insight into the universality of the heavy r-process. The formation of a long-lived or stable remnant NS \cite{Metzger2014b,Siegel2016a,Siegel2016b}, for instance, could lead to an essentially lanthanide-free (blue) merger event. In parallel, experimental nuclear facilities such as TRIUMF and FRIB will experimentally characterize a large number of r-process nuclei over the next decade. Together with advanced hydrodynamic evolution codes this will enable precision astrophysical modeling of various r-process sites.

Observations with future sensitive NIR telescopes, such as the James Webb Space Telescope, may allow us to identify unique signatures of heavy r-process nuclei in kilonova emission to directly infer the composition of merger ejecta. Applied to future GRB supernovae at late times, one may be able to directly probe r-process production by collapsars. Together with more precise merger rates and ejecta masses from a statistical ensemble of events, the relative contribution of NS mergers to collapsars can be more tightly constrained. This has important consequences for r-process enrichment in metal-poor halo stars, dwarf galaxies, globular clusters, the Milky Way disk, and for the dynamical assembly of the Milky Way halo.

\section*{Acknowledgments}
The author thanks K.~Hotokezaka, P.~Beniamini, and D. Radice for sharing data in \cite{Hotokezaka+15,Beniamini2016a,Radice2018b}. D.~M.~S. also thanks B.~Metzger and J.~Barnes, with whom some of the results summarized and discussed here have been obtained. The author thanks R.~Sch\"onrich for discussions and for pointing out the problem of r-process versus $\alpha$-element production. This work was performed in part at the Aspen Center for Physics, which is supported by National Science Foundation grant PHY-1607611; the author thanks the participants of the Aspen conference `Astrophysics with Gravitational-Wave Populations' for stimulating discussions and the Aspen Center for hospitality. Resources supporting this work were provided by the NASA High-End Computing (HEC) Program through the NASA Advanced Supercomputing (NAS) Division at Ames Research Center. Support for this work was provided by the National Aeronautics and Space Administration through Einstein Postdoctoral Fellowship Award Number PF6-170159 issued by the Chandra X-ray Observatory Center, which is operated by the Smithsonian Astrophysical Observatory for and on behalf of the National Aeronautics Space Administration under contract NAS8-03060. Research at Perimeter Institute is supported in part by the Government of Canada through the Department of Innovation, Science and Economic Development Canada and by the Province of Ontario through the Ministry of Economic Development, Job Creation and Trade.

\bibliographystyle{epj}

\begin{thebibliography}{214}

\bibitem{Suess1956}
H.E. {Suess}, H.C. {Urey}, Rev.~Mod.~Phys.~\textbf{28}, 53 (1956)

\bibitem{Burbidge1957}
E.M. {Burbidge}, G.R. {Burbidge}, W.A. {Fowler}, F.~{Hoyle}, Rev.~Mod.~Phys.~\textbf{29},
  547 (1957)

\bibitem{Cameron1957}
A.G.W. {Cameron}, Astron.~J.~\textbf{62}, 9 (1957)

\bibitem{Cameron57}
A.G.W. {Cameron}, Publ. Astron. Soc. Pac. \textbf{69}, 201 (1957)

\bibitem{Thielemann2011}
F.K. {Thielemann}, A.~{Arcones}, R.~{K{\"a}ppeli}, M.~{Liebend{\"o}rfer},
  T.~{Rauscher}, C.~{Winteler}, C.~{Fr{\"o}hlich}, I.~{Dillmann}, T.~{Fischer},
  G.~{Martinez-Pinedo} et~al., Prog. Part. Nucl. Phys. \textbf{66}, 346 (2011)

\bibitem{Qian&Woosley96}
Y.~{Qian}, S.E. {Woosley}, Astrophys. J. \textbf{471}, 331 (1996),
  \texttt{arXiv:astro-ph/9611094}

\bibitem{Thompson+01}
T.A. {Thompson}, A.~{Burrows}, B.S. {Meyer}, Astrophys. J. \textbf{562}, 887
  (2001), \texttt{arXiv:astro-ph/0105004}

\bibitem{Arcones+07}
A.~{Arcones}, H.T. {Janka}, L.~{Scheck}, Astron. Astrophys. \textbf{467}, 1227
  (2007), \texttt{astro-ph/0612582}

\bibitem{Roberts+12}
L.F. {Roberts}, S.~{Reddy}, G.~{Shen}, Phys. Rev. C \textbf{86}, 065803 (2012),
  \texttt{1205.4066}

\bibitem{MartinezPinedo+12}
G.~{Mart{\'{\i}}nez-Pinedo}, T.~{Fischer}, A.~{Lohs}, L.~{Huther}, Phys. Rev.
  Lett. \textbf{109}, 251104 (2012), \texttt{1205.2793}

\bibitem{Wallner+15}
A.~{Wallner}, T.~{Faestermann}, J.~{Feige}, C.~{Feldstein}, K.~{Knie},
  G.~{Korschinek}, W.~{Kutschera}, A.~{Ofan}, M.~{Paul}, F.~{Quinto} et~al.,
  Nature Commun. \textbf{6}, 5956 (2015), \texttt{1509.08054}

\bibitem{Hotokezaka+15}
K.~{Hotokezaka}, T.~{Piran}, M.~{Paul}, Nature Phys. \textbf{11}, 1042 (2015),
  \texttt{1510.00711}

\bibitem{Ji2016}
A.P. {Ji}, A.~{Frebel}, A.~{Chiti}, J.D. {Simon}, \nat \textbf{531}, 610
  (2016), \texttt{1512.01558}

\bibitem{Lattimer1974}
J.M. {Lattimer}, D.N. {Schramm}, \apjl \textbf{192}, L145 (1974)

\bibitem{Lattimer1976}
J.M. {Lattimer}, D.N. {Schramm}, \apj \textbf{210}, 549 (1976)

\bibitem{Hulse1975}
R.A. {Hulse}, J.H. {Taylor}, \apjl \textbf{195}, L51 (1975)

\bibitem{Symbalisty&Schramm82}
E.~{Symbalisty}, D.N. {Schramm}, Astrophys. J. Lett. \textbf{22}, 143 (1982)

\bibitem{Eichler1989}
D.~{Eichler}, M.~{Livio}, T.~{Piran}, D.N. {Schramm}, \nat \textbf{340}, 126
  (1989)

\bibitem{Meyer89}
B.S. {Meyer}, Astrophys. J. \textbf{343}, 254 (1989)

\bibitem{Davies+94}
M.B. {Davies}, W.~{Benz}, T.~{Piran}, F.K. {Thielemann}, Astrophys. J.
  \textbf{431}, 742 (1994), \texttt{astro-ph/9401032}

\bibitem{Ruffert1996b}
M.~{Ruffert}, H.T. {Janka}, G.~{Schaefer}, \aap \textbf{311}, 532 (1996),
  \texttt{astro-ph/9509006}

\bibitem{Ruffert1997}
M.~{Ruffert}, H.T. {Janka}, K.~{Takahashi}, G.~{Schaefer}, \aap \textbf{319},
  122 (1997), \texttt{astro-ph/9606181}

\bibitem{Rosswog1999}
S.~{Rosswog}, M.~{Liebend{\"o}rfer}, F.K. {Thielemann}, M.B. {Davies},
  W.~{Benz}, T.~{Piran}, \aap \textbf{341}, 499 (1999),
  \texttt{astro-ph/9811367}

\bibitem{Freiburghaus1999}
C.~{Freiburghaus}, S.~{Rosswog}, F.K. {Thielemann}, \apjl \textbf{525}, L121
  (1999)

\bibitem{LIGO+17DISCOVERY}
B.P. {Abbott}, R.~{Abbott}, T.D. {Abbott}, F.~{Acernese}, K.~{Ackley},
  C.~{Adams}, T.~{Adams}, P.~{Addesso}, R.X. {Adhikari}, V.B. {Adya} et~al.,
  \prl \textbf{119}, 161101 (2017), \texttt{1710.05832}

\bibitem{Coulter2017}
D.A. {Coulter}, R.J. {Foley}, C.D. {Kilpatrick}, M.R. {Drout}, A.L. {Piro},
  B.J. {Shappee}, M.R. {Siebert}, J.D. {Simon}, N.~{Ulloa}, D.~{Kasen} et~al.,
  Science \textbf{358}, 1556 (2017), \texttt{1710.05452}

\bibitem{Soares-Santos+17}
M.~{Soares-Santos} et~al., \apjl \textbf{848}, L16 (2017), \texttt{1710.05459}

\bibitem{LIGO+17CAPSTONE}
B.P. {Abbott}, R.~{Abbott}, T.D. {Abbott}, F.~{Acernese}, K.~{Ackley},
  C.~{Adams}, T.~{Adams}, P.~{Addesso}, R.X. {Adhikari}, V.B. {Adya} et~al.,
  \apjl \textbf{848}, L12 (2017), \texttt{1710.05833}

\bibitem{Li1998}
L.X. {Li}, B.~{Paczy{\'n}ski}, \apjl \textbf{507}, L59 (1998),
  \texttt{astro-ph/9807272}

\bibitem{Kulkarni2005}
S.R. {Kulkarni}, ArXiv Astrophysics e-prints  (2005), \texttt{astro-ph/0510256}

\bibitem{Metzger2010}
B.D. {Metzger}, G.~{Mart{\'{\i}}nez-Pinedo}, S.~{Darbha}, E.~{Quataert},
  A.~{Arcones}, D.~{Kasen}, R.~{Thomas}, P.~{Nugent}, I.V. {Panov}, N.T.
  {Zinner}, \mnras \textbf{406}, 2650 (2010), \texttt{1001.5029}

\bibitem{Metzger17}
B.D. {Metzger}, Living Reviews in Relativity \textbf{20}, 3 (2017),
  \texttt{1610.09381}

\bibitem{Nicholl2017}
M.~{Nicholl}, E.~{Berger}, D.~{Kasen}, B.D. {Metzger}, J.~{Elias},
  C.~{Brice{\~n}o}, K.D. {Alexander}, P.K. {Blanchard}, R.~{Chornock}, P.S.
  {Cowperthwaite} et~al., \apjl \textbf{848}, L18 (2017), \texttt{1710.05456}

\bibitem{Cowperthwaite2017}
P.S. {Cowperthwaite}, E.~{Berger}, V.A. {Villar}, B.D. {Metzger}, M.~{Nicholl},
  R.~{Chornock}, P.K. {Blanchard}, W.~{Fong}, R.~{Margutti}, M.~{Soares-Santos}
  et~al., \apjl \textbf{848}, L17 (2017), \texttt{1710.05840}

\bibitem{Kasen2017}
D.~{Kasen}, B.~{Metzger}, J.~{Barnes}, E.~{Quataert}, E.~{Ramirez-Ruiz}, \nat
  \textbf{551}, 80 (2017), \texttt{1710.05463}

\bibitem{Kasliwal2017}
M.M. {Kasliwal}, E.~{Nakar}, L.P. {Singer}, D.L. {Kaplan}, D.O. {Cook}, A.~{Van
  Sistine}, R.M. {Lau}, C.~{Fremling}, O.~{Gottlieb}, J.E. {Jencson} et~al.,
  Science \textbf{358}, 1559 (2017), \texttt{1710.05436}

\bibitem{Drout2017}
M.R. {Drout}, A.L. {Piro}, B.J. {Shappee}, C.D. {Kilpatrick}, J.D. {Simon},
  C.~{Contreras}, D.A. {Coulter}, R.J. {Foley}, M.R. {Siebert}, N.~{Morrell}
  et~al., Science \textbf{358}, 1570 (2017), \texttt{1710.05443}

\bibitem{Smartt2017}
S.J. {Smartt}, T.W. {Chen}, A.~{Jerkstrand}, M.~{Coughlin}, E.~{Kankare}, S.A.
  {Sim}, M.~{Fraser}, C.~{Inserra}, K.~{Maguire}, K.C. {Chambers} et~al., \nat
  \textbf{551}, 75 (2017), \texttt{1710.05841}

\bibitem{Troja2017}
E.~{Troja}, L.~{Piro}, H.~{van Eerten}, R.T. {Wollaeger}, M.~{Im}, O.D. {Fox},
  N.R. {Butler}, S.B. {Cenko}, T.~{Sakamoto}, C.L. {Fryer} et~al., \nat
  \textbf{551}, 71 (2017), \texttt{1710.05433}

\bibitem{Kilpatrick2017}
C.D. {Kilpatrick}, R.J. {Foley}, D.~{Kasen}, A.~{Murguia-Berthier},
  E.~{Ramirez-Ruiz}, D.A. {Coulter}, M.R. {Drout}, A.L. {Piro}, B.J. {Shappee},
  K.~{Boutsia} et~al., Science \textbf{358}, 1583 (2017), \texttt{1710.05434}

\bibitem{Evans2017}
P.A. {Evans}, S.B. {Cenko}, J.A. {Kennea}, S.W.K. {Emery}, N.P.M. {Kuin},
  O.~{Korobkin}, R.T. {Wollaeger}, C.L. {Fryer}, K.K. {Madsen}, F.A. {Harrison}
  et~al., Science \textbf{358}, 1565 (2017), \texttt{1710.05437}

\bibitem{Pian2017}
E.~{Pian}, P.~{D'Avanzo}, S.~{Benetti}, M.~{Branchesi}, E.~{Brocato},
  S.~{Campana}, E.~{Cappellaro}, S.~{Covino}, V.~{D'Elia}, J.P.U. {Fynbo}
  et~al., \nat \textbf{551}, 67 (2017), \texttt{1710.05858}

\bibitem{McCully2017}
C.~{McCully}, D.~{Hiramatsu}, D.A. {Howell}, G.~{Hosseinzadeh}, I.~{Arcavi},
  D.~{Kasen}, J.~{Barnes}, M.M. {Shara}, T.B. {Williams}, P.~{V{\"a}is{\"a}nen}
  et~al., \apjl \textbf{848}, L32 (2017), \texttt{1710.05853}

\bibitem{Arcavi2017}
I.~{Arcavi}, G.~{Hosseinzadeh}, D.A. {Howell}, C.~{McCully}, D.~{Poznanski},
  D.~{Kasen}, J.~{Barnes}, M.~{Zaltzman}, S.~{Vasylyev}, D.~{Maoz} et~al., \nat
  \textbf{551}, 64 (2017), \texttt{1710.05843}

\bibitem{Andreoni2017}
I.~{Andreoni}, K.~{Ackley}, J.~{Cooke}, A.~{Acharyya}, J.R. {Allison}, G.E.
  {Anderson}, M.C.B. {Ashley}, D.~{Baade}, M.~{Bailes}, K.~{Bannister} et~al.,
  \pasa \textbf{34}, e069 (2017), \texttt{1710.05846}

\bibitem{Diaz2017}
M.C. {D{\'{\i}}az}, L.M. {Macri}, D.~{Garcia Lambas}, C.~{Mendes de Oliveira},
  J.L. {Nilo Castell{\'o}n}, T.~{Ribeiro}, B.~{S{\'a}nchez}, W.~{Schoenell},
  L.R. {Abramo}, S.~{Akras} et~al., \apjl \textbf{848}, L29 (2017),
  \texttt{1710.05844}

\bibitem{Lipunov2017}
V.M. {Lipunov}, E.~{Gorbovskoy}, V.G. {Kornilov}, N.~{.~Tyurina},
  P.~{Balanutsa}, A.~{Kuznetsov}, D.~{Vlasenko}, D.~{Kuvshinov}, I.~{Gorbunov},
  D.A.H. {Buckley} et~al., \apjl \textbf{850}, L1 (2017), \texttt{1710.05461}

\bibitem{Valenti2017}
S.~{Valenti}, {David}, J.~{Sand}, S.~{Yang}, E.~{Cappellaro}, L.~{Tartaglia},
  A.~{Corsi}, S.W. {Jha}, D.E. {Reichart}, J.~{Haislip} et~al., \apjl
  \textbf{848}, L24 (2017), \texttt{1710.05854}

\bibitem{Tanvir2017}
N.R. {Tanvir}, A.J. {Levan}, C.~{Gonz{\'a}lez-Fern{\'a}ndez}, O.~{Korobkin},
  I.~{Mandel}, S.~{Rosswog}, J.~{Hjorth}, P.~{D'Avanzo}, A.S. {Fruchter}, C.L.
  {Fryer} et~al., \apjl \textbf{848}, L27 (2017), \texttt{1710.05455}

\bibitem{Utsumi2017}
Y.~{Utsumi}, M.~{Tanaka}, N.~{Tominaga}, M.~{Yoshida}, S.~{Barway},
  T.~{Nagayama}, T.~{Zenko}, K.~{Aoki}, T.~{Fujiyoshi}, H.~{Furusawa} et~al.,
  \pasj \textbf{69}, 101 (2017), \texttt{1710.05848}

\bibitem{Villar2017}
V.A. {Villar}, J.~{Guillochon}, E.~{Berger}, B.D. {Metzger}, P.S.
  {Cowperthwaite}, M.~{Nicholl}, K.D. {Alexander}, P.K. {Blanchard},
  R.~{Chornock}, T.~{Eftekhari} et~al., \apjl \textbf{851}, L21 (2017),
  \texttt{1710.11576}

\bibitem{Siebert2017}
M.R. {Siebert}, R.J. {Foley}, M.R. {Drout}, C.D. {Kilpatrick}, B.J. {Shappee},
  D.A. {Coulter}, D.~{Kasen}, B.F. {Madore}, A.~{Murguia-Berthier}, Y.C. {Pan}
  et~al., \apjl \textbf{848}, L26 (2017), \texttt{1710.05440}

\bibitem{Kasen2013}
D.~{Kasen}, N.R. {Badnell}, J.~{Barnes}, \apj \textbf{774}, 25 (2013),
  \texttt{1303.5788}

\bibitem{Barnes2013}
J.~{Barnes}, D.~{Kasen}, \apj \textbf{775}, 18 (2013), \texttt{1303.5787}

\bibitem{Tanaka2013}
M.~{Tanaka}, K.~{Hotokezaka}, \apj \textbf{775}, 113 (2013), \texttt{1306.3742}

\bibitem{Chornock2017}
R.~{Chornock}, E.~{Berger}, D.~{Kasen}, P.S. {Cowperthwaite}, M.~{Nicholl},
  V.A. {Villar}, K.D. {Alexander}, P.K. {Blanchard}, T.~{Eftekhari}, W.~{Fong}
  et~al., \apjl \textbf{848}, L19 (2017), \texttt{1710.05454}

\bibitem{Metzger2017b}
B.D. {Metzger}, arXiv e-prints  (2017), \texttt{1710.05931}

\bibitem{Arnett1979}
W.D. {Arnett}, \apjl \textbf{230}, L37 (1979)

\bibitem{Arnett1982}
W.D. {Arnett}, \apj \textbf{253}, 785 (1982)

\bibitem{Perego+17}
A.~{Perego}, D.~{Radice}, S.~{Bernuzzi}, \apjl \textbf{850}, L37 (2017),
  \texttt{1711.03982}

\bibitem{Coughlin2018a}
M.W. {Coughlin}, T.~{Dietrich}, Z.~{Doctor}, D.~{Kasen}, S.~{Coughlin},
  A.~{Jerkstrand}, G.~{Leloudas}, O.~{McBrien}, B.D. {Metzger},
  R.~{O'Shaughnessy} et~al., \mnras \textbf{480}, 3871 (2018),
  \texttt{1805.09371}

\bibitem{Tanaka2017}
M.~{Tanaka}, Y.~{Utsumi}, P.A. {Mazzali}, N.~{Tominaga}, M.~{Yoshida},
  Y.~{Sekiguchi}, T.~{Morokuma}, K.~{Motohara}, K.~{Ohta}, K.S. {Kawabata}
  et~al., \pasj \textbf{69}, 102 (2017), \texttt{1710.05850}

\bibitem{Waxman2018}
E.~{Waxman}, E.O. {Ofek}, D.~{Kushnir}, A.~{Gal-Yam}, \mnras \textbf{481}, 3423
  (2018), \texttt{1711.09638}

\bibitem{Kawaguchi2018}
K.~{Kawaguchi}, M.~{Shibata}, M.~{Tanaka}, \apjl \textbf{865}, L21 (2018),
  \texttt{1806.04088}

\bibitem{Lippuner2015}
J.~{Lippuner}, L.F. {Roberts}, \apj \textbf{815}, 82 (2015),
  \texttt{1508.03133}

\bibitem{Tanaka+18}
M.~{Tanaka}, D.~{Kato}, G.~{Gaigalas}, P.~{Rynkun}, L.~{Rad{\v z}i{\=
  u}t{\.e}}, S.~{Wanajo}, Y.~{Sekiguchi}, N.~{Nakamura}, H.~{Tanuma},
  I.~{Murakami} et~al., \apj \textbf{852}, 109 (2018), \texttt{1708.09101}

\bibitem{Sekiguchi2016}
Y.~{Sekiguchi}, K.~{Kiuchi}, K.~{Kyutoku}, M.~{Shibata}, K.~{Taniguchi}, \prd
  \textbf{93}, 124046 (2016), \texttt{1603.01918}

\bibitem{Radice2018b}
D.~{Radice}, A.~{Perego}, K.~{Hotokezaka}, S.A. {Fromm}, S.~{Bernuzzi}, L.F.
  {Roberts}, ArXiv e-prints  (2018), \texttt{1809.11161}

\bibitem{Siegel2018a}
D.M. {Siegel}, B.D. {Metzger}, \apj \textbf{858}, 52 (2018)

\bibitem{Fernandez2019}
R.~{Fern{\'a}ndez}, A.~{Tchekhovskoy}, E.~{Quataert}, F.~{Foucart}, D.~{Kasen},
  \mnras \textbf{482}, 3373 (2019), \texttt{1808.00461}

\bibitem{Miller2019}
J.M. {Miller}, B.R. {Ryan}, J.C. {Dolence}, A.~{Burrows}, C.J. {Fontes}, C.L.
  {Fryer}, O.~{Korobkin}, J.~{Lippuner}, M.R. {Mumpower}, R.T. {Wollaeger},
  arXiv e-prints arXiv:1905.07477 (2019), \texttt{1905.07477}

\bibitem{Metzger2018a}
B.D. {Metzger}, T.A. {Thompson}, E.~{Quataert}, \apj \textbf{856}, 101 (2018),
  \texttt{1801.04286}

\bibitem{Kasen+15}
D.~{Kasen}, R.~{Fern{\'a}ndez}, B.D. {Metzger}, Mon. Not. R. Astron. Soc.
  \textbf{450}, 1777 (2015), \texttt{1411.3726}

\bibitem{Wollaeger2018}
R.T. {Wollaeger}, O.~{Korobkin}, C.J. {Fontes}, S.K. {Rosswog}, W.P. {Even},
  C.L. {Fryer}, J.~{Sollerman}, A.L. {Hungerford}, D.R. {van Rossum}, A.B.
  {Wollaber}, \mnras \textbf{478}, 3298 (2018), \texttt{1705.07084}

\bibitem{Barnes+16}
J.~{Barnes}, D.~{Kasen}, M.R. {Wu}, G.~{Mart{\'{\i}}nez-Pinedo}, Astrophys. J.
  \textbf{829}, 110 (2016), \texttt{1605.07218}

\bibitem{Korobkin2012}
O.~{Korobkin}, S.~{Rosswog}, A.~{Arcones}, C.~{Winteler}, \mnras \textbf{426},
  1940 (2012), \texttt{1206.2379}

\bibitem{Rosswog2017a}
S.~{Rosswog}, U.~{Feindt}, O.~{Korobkin}, M.R. {Wu}, J.~{Sollerman},
  A.~{Goobar}, G.~{Martinez-Pinedo}, Classical and Quantum Gravity \textbf{34},
  104001 (2017), \texttt{1611.09822}

\bibitem{Wu2019a}
M.R. {Wu}, J.~{Barnes}, G.~{Mart{\'{\i}}nez-Pinedo}, B.D. {Metzger}, \prl
  \textbf{122}, 062701 (2019), \texttt{1808.10459}

\bibitem{Oechslin2007}
R.~{Oechslin}, H.T. {Janka}, A.~{Marek}, \aap \textbf{467}, 395 (2007),
  \texttt{astro-ph/0611047}

\bibitem{Hotokezaka2013a}
K.~{Hotokezaka}, K.~{Kyutoku}, M.~{Tanaka}, K.~{Kiuchi}, Y.~{Sekiguchi},
  M.~{Shibata}, S.~{Wanajo}, \apjl \textbf{778}, L16 (2013), \texttt{1310.1623}

\bibitem{Dessart2009}
L.~{Dessart}, C.D. {Ott}, A.~{Burrows}, S.~{Rosswog}, E.~{Livne}, \apj
  \textbf{690}, 1681 (2009), \texttt{0806.4380}

\bibitem{Siegel2014a}
D.M. {Siegel}, R.~{Ciolfi}, L.~{Rezzolla}, \apjl \textbf{785}, L6 (2014),
  \texttt{1401.4544}

\bibitem{Ciolfi2017a}
R.~{Ciolfi}, W.~{Kastaun}, B.~{Giacomazzo}, A.~{Endrizzi}, D.M. {Siegel},
  R.~{Perna}, \prd \textbf{95}, 063016 (2017), \texttt{1701.08738}

\bibitem{Ciolfi2019}
R.~Ciolfi, W.~Kastaun, J.V. Kalinani, B.~Giacomazzo, \prd \textbf{100}, 023005
  (2019)

\bibitem{Fernandez2013}
R.~{Fern{\'a}ndez}, B.D. {Metzger}, \mnras \textbf{435}, 502 (2013),
  \texttt{1304.6720}

\bibitem{Just2015a}
O.~{Just}, A.~{Bauswein}, R.A. {Pulpillo}, S.~{Goriely}, H.T. {Janka}, \mnras
  \textbf{448}, 541 (2015), \texttt{1406.2687}

\bibitem{Siegel2017a}
D.M. {Siegel}, B.D. {Metzger}, \prl \textbf{119}, 231102 (2017),
  \texttt{1705.05473}

\bibitem{Bauswein2013a}
A.~{Bauswein}, S.~{Goriely}, H.T. {Janka}, \apj \textbf{773}, 78 (2013),
  \texttt{1302.6530}

\bibitem{Hotokezaka2013b}
K.~{Hotokezaka}, K.~{Kiuchi}, K.~{Kyutoku}, H.~{Okawa}, Y.i. {Sekiguchi},
  M.~{Shibata}, K.~{Taniguchi}, \prd \textbf{87}, 024001 (2013),
  \texttt{1212.0905}

\bibitem{Sekiguchi2015}
Y.~{Sekiguchi}, K.~{Kiuchi}, K.~{Kyutoku}, M.~{Shibata}, Phys. Rev. D
  \textbf{91}, 064059 (2015), \texttt{1502.06660}

\bibitem{Kyutoku2014}
K.~{Kyutoku}, K.~{Ioka}, M.~{Shibata}, \mnras \textbf{437}, L6 (2014),
  \texttt{1209.5747}

\bibitem{Dietrich2015a}
T.~{Dietrich}, S.~{Bernuzzi}, M.~{Ujevic}, B.~{Br{\"u}gmann}, \prd \textbf{91},
  124041 (2015), \texttt{1504.01266}

\bibitem{Kastaun2016a}
W.~{Kastaun}, R.~{Ciolfi}, B.~{Giacomazzo}, \prd \textbf{94}, 044060 (2016),
  \texttt{1607.02186}

\bibitem{Hotokezaka2018b}
K.~{Hotokezaka}, K.~{Kiuchi}, M.~{Shibata}, E.~{Nakar}, T.~{Piran}, \apj
  \textbf{867}, 95 (2018), \texttt{1803.00599}

\bibitem{Metzger+15}
B.D. {Metzger}, A.~{Bauswein}, S.~{Goriely}, D.~{Kasen}, Mon. Not. R. Astron.
  Soc. \textbf{446}, 1115 (2015), \texttt{1409.0544}

\bibitem{Radice2016}
D.~{Radice}, F.~{Galeazzi}, J.~{Lippuner}, L.F. {Roberts}, C.D. {Ott},
  L.~{Rezzolla}, Mon. Not. R. Astron. Soc.  (2016), \texttt{1601.02426}

\bibitem{Lehner2016}
L.~{Lehner}, S.L. {Liebling}, C.~{Palenzuela}, O.L. {Caballero}, E.~{O'Connor},
  M.~{Anderson}, D.~{Neilsen}, Classical and Quantum Gravity \textbf{33},
  184002 (2016), \texttt{1603.00501}

\bibitem{Palenzuela2015}
C.~{Palenzuela}, S.L. {Liebling}, D.~{Neilsen}, L.~{Lehner}, O.L. {Caballero},
  E.~{O'Connor}, M.~{Anderson}, \prd \textbf{92}, 044045 (2015),
  \texttt{1505.01607}

\bibitem{Radice2018a}
D.~{Radice}, A.~{Perego}, F.~{Zappa}, S.~{Bernuzzi}, \apjl \textbf{852}, L29
  (2018), \texttt{1711.03647}

\bibitem{Burgay03}
M.~Burgay, N.~D'Amico, A.~Possenti, R.~Manchester, A.~Lyne, B.~Joshi,
  M.~McLaughlin, M.~Kramer, J.~Sarkissian, F.~Camilo et~al., Nature
  \textbf{426}, 531 (2003)

\bibitem{Kastaun2015a}
W.~{Kastaun}, F.~{Galeazzi}, \prd \textbf{91}, 064027 (2015),
  \texttt{1411.7975}

\bibitem{Dietrich2017a}
T.~{Dietrich}, M.~{Ujevic}, W.~{Tichy}, S.~{Bernuzzi}, B.~{Br{\"u}gmann}, \prd
  \textbf{95}, 024029 (2017), \texttt{1607.06636}

\bibitem{East2019}
W.E. {East}, V.~{Paschalidis}, F.~{Pretorius}, A.~{Tsokaros}, arXiv e-prints
  arXiv:1906.05288 (2019), \texttt{1906.05288}

\bibitem{Most2019}
E.R. {Most}, L.J. {Papenfort}, A.~{Tsokaros}, L.~{Rezzolla}, arXiv e-prints
  arXiv:1904.04220 (2019), \texttt{1904.04220}

\bibitem{Dietrich2018a}
T.~{Dietrich}, S.~{Bernuzzi}, B.~{Br{\"u}gmann}, M.~{Ujevic}, W.~{Tichy}, \prd
  \textbf{97}, 064002 (2018), \texttt{1712.02992}

\bibitem{Gold2012}
R.~{Gold}, S.~{Bernuzzi}, M.~{Thierfelder}, B.~{Br{\"u}gmann}, F.~{Pretorius},
  \prd \textbf{86}, 121501 (2012), \texttt{1109.5128}

\bibitem{East2016}
W.E. {East}, V.~{Paschalidis}, F.~{Pretorius}, S.L. {Shapiro}, \prd
  \textbf{93}, 024011 (2016), \texttt{1511.01093}

\bibitem{Chaurasia2018}
S.V. {Chaurasia}, T.~{Dietrich}, N.K. {Johnson-McDaniel}, M.~{Ujevic},
  W.~{Tichy}, B.~{Br{\"u}gmann}, \prd \textbf{98}, 104005 (2018)

\bibitem{Ozel+16}
F.~{{\"O}zel}, D.~{Psaltis}, T.~{G{\"u}ver}, G.~{Baym}, C.~{Heinke},
  S.~{Guillot}, \apj \textbf{820}, 28 (2016), \texttt{1505.05155}

\bibitem{Hotokezaka2011}
K.~{Hotokezaka}, K.~{Kyutoku}, H.~{Okawa}, M.~{Shibata}, K.~{Kiuchi}, \prd
  \textbf{83}, 124008 (2011), \texttt{1105.4370}

\bibitem{Bauswein2013b}
A.~{Bauswein}, T.W. {Baumgarte}, H.T. {Janka}, \prl \textbf{111}, 131101
  (2013), \texttt{1307.5191}

\bibitem{Shapiro2000}
S.L. {Shapiro}, \apj \textbf{544}, 397 (2000)

\bibitem{Duez2006a}
M.D. {Duez}, Y.T. {Liu}, S.L. {Shapiro}, M.~{Shibata}, B.C. {Stephens}, \prl
  \textbf{96}, 031101 (2006), \texttt{astro-ph/0510653}

\bibitem{Siegel2013}
D.M. {Siegel}, R.~{Ciolfi}, A.I. {Harte}, L.~{Rezzolla}, \prd \textbf{87},
  121302 (2013), \texttt{1302.4368}

\bibitem{Kaplan2014}
J.D. {Kaplan}, C.D. {Ott}, E.P. {O'Connor}, K.~{Kiuchi}, L.~{Roberts},
  M.~{Duez}, \apj \textbf{790}, 19 (2014), \texttt{1306.4034}

\bibitem{Fujibayashi+18}
S.~{Fujibayashi}, K.~{Kiuchi}, N.~{Nishimura}, Y.~{Sekiguchi}, M.~{Shibata},
  \apj \textbf{860}, 64 (2018), \texttt{1711.02093}

\bibitem{Radice2018c}
D.~{Radice}, A.~{Perego}, K.~{Hotokezaka}, S.~{Bernuzzi}, S.A. {Fromm}, L.F.
  {Roberts}, \apjl \textbf{869}, L35 (2018), \texttt{1809.11163}

\bibitem{Shibata2006a}
M.~{Shibata}, K.~{Taniguchi}, \prd \textbf{73}, 064027 (2006),
  \texttt{astro-ph/0603145}

\bibitem{Wu2016}
M.R. {Wu}, R.~{Fern{\'a}ndez}, G.~{Mart{\'{\i}}nez-Pinedo}, B.D. {Metzger},
  \mnras \textbf{463}, 2323 (2016), \texttt{1607.05290}

\bibitem{Metzger2014c}
B.D. {Metzger}, R.~{Fern{\'a}ndez}, \mnras \textbf{441}, 3444 (2014),
  \texttt{1402.4803}

\bibitem{Lippuner2017a}
J.~{Lippuner}, R.~{Fern{\'a}ndez}, L.F. {Roberts}, F.~{Foucart}, D.~{Kasen},
  B.D. {Metzger}, C.D. {Ott}, \mnras \textbf{472}, 904 (2017),
  \texttt{1703.06216}

\bibitem{Fahlman2018}
S.~{Fahlman}, R.~{Fern{\'a}ndez}, \apjl \textbf{869}, L3 (2018),
  \texttt{1811.08906}

\bibitem{Nedora2019}
V.~{Nedora}, S.~{Bernuzzi}, D.~{Radice}, A.~{Perego}, A.~{Endrizzi},
  N.~{Ortiz}, arXiv e-prints arXiv:1907.04872 (2019), \texttt{1907.04872}

\bibitem{Tauris2017}
T.M. {Tauris}, M.~{Kramer}, P.C.C. {Freire}, N.~{Wex}, H.T. {Janka},
  N.~{Langer}, P.~{Podsiadlowski}, E.~{Bozzo}, S.~{Chaty}, M.U. {Kruckow}
  et~al., \apj \textbf{846}, 170 (2017), \texttt{1706.09438}

\bibitem{Ozel2010a}
F.~{{\"O}zel}, D.~{Psaltis}, R.~{Narayan}, J.E. {McClintock}, \apj
  \textbf{725}, 1918 (2010), \texttt{1006.2834}

\bibitem{Kreidberg2012}
L.~{Kreidberg}, C.D. {Bailyn}, W.M. {Farr}, V.~{Kalogera}, \apj \textbf{757},
  36 (2012), \texttt{1205.1805}

\bibitem{Yang2018}
H.~{Yang}, W.E. {East}, L.~{Lehner}, \apj \textbf{856}, 110 (2018),
  \texttt{1710.05891}

\bibitem{Foucart2019}
F.~{Foucart}, M.D. {Duez}, L.E. {Kidder}, S.M. {Nissanke}, H.P. {Pfeiffer},
  M.A. {Scheel}, \prd \textbf{99}, 103025 (2019), \texttt{1903.09166}

\bibitem{Li2011}
W.~{Li}, R.~{Chornock}, J.~{Leaman}, A.V. {Filippenko}, D.~{Poznanski},
  X.~{Wang}, M.~{Ganeshalingam}, F.~{Mannucci}, \mnras \textbf{412}, 1473
  (2011), \texttt{1006.4613}

\bibitem{Beniamini2016a}
P.~{Beniamini}, K.~{Hotokezaka}, T.~{Piran}, \apj \textbf{832}, 149 (2016),
  \texttt{1608.08650}

\bibitem{Macias2018}
P.~{Macias}, E.~{Ramirez-Ruiz}, \apj \textbf{860}, 89 (2018),
  \texttt{1609.04826}

\bibitem{Arnould2007}
M.~{Arnould}, S.~{Goriely}, K.~{Takahashi}, \physrep \textbf{450}, 97 (2007),
  \texttt{0705.4512}

\bibitem{Kopparapu2008}
R.K. {Kopparapu}, C.~{Hanna}, V.~{Kalogera}, R.~{O'Shaughnessy},
  G.~{Gonz{\'a}lez}, P.R. {Brady}, S.~{Fairhurst}, \apj \textbf{675}, 1459
  (2008), \texttt{0706.1283}

\bibitem{Madau2017}
P.~{Madau}, T.~{Fragos}, \apj \textbf{840}, 39 (2017), \texttt{1606.07887}

\bibitem{Siegel2019a}
D.M. {Siegel}, J.~{Barnes}, B.D. {Metzger}, \nat \textbf{569}, 241 (2019),
  \texttt{1810.00098}

\bibitem{McMillan2011}
P.J. {McMillan}, \mnras \textbf{414}, 2446 (2011), \texttt{1102.4340}

\bibitem{Heckman2000}
T.M. {Heckman}, M.D. {Lehnert}, D.K. {Strickland}, L.~{Armus}, \apjs
  \textbf{129}, 493 (2000), \texttt{astro-ph/0002526}

\bibitem{Murray2005}
N.~{Murray}, E.~{Quataert}, T.A. {Thompson}, \apj \textbf{618}, 569 (2005),
  \texttt{astro-ph/0406070}

\bibitem{Chisholm2017}
J.~{Chisholm}, C.A. {Tremonti}, C.~{Leitherer}, Y.~{Chen}, \mnras \textbf{469},
  4831 (2017), \texttt{1702.07351}

\bibitem{Cote2016a}
B.~{C{\^o}t{\'e}}, C.~{West}, A.~{Heger}, C.~{Ritter}, B.W. {O'Shea},
  F.~{Herwig}, C.~{Travaglio}, S.~{Bisterzo}, \mnras \textbf{463}, 3755 (2016),
  \texttt{1602.04824}

\bibitem{Hotokezaka2018a}
K.~{Hotokezaka}, P.~{Beniamini}, T.~{Piran}, Int. J. Mod. Phys. D \textbf{27},
  1842005 (2018), \texttt{1801.01141}

\bibitem{Turner2007}
G.~{Turner}, A.~{Busfield}, S.A. {Crowther}, M.~{Harrison}, S.J. {Mojzsis},
  J.~{Gilmour}, Earth and Planetary Science Letters \textbf{261}, 491 (2007)

\bibitem{Hansen2017}
T.T. {Hansen}, J.D. {Simon}, J.L. {Marshall}, T.S. {Li}, D.~{Carollo}, D.L.
  {DePoy}, D.Q. {Nagasawa}, R.A. {Bernstein}, A.~{Drlica-Wagner}, F.B.
  {Abdalla} et~al., \apj \textbf{838}, 44 (2017), \texttt{1702.07430}

\bibitem{Berger2014}
E.~{Berger}, \araa \textbf{52}, 43 (2014), \texttt{1311.2603}

\bibitem{Cote2018a}
B.~{C{\^o}t{\'e}}, C.L. {Fryer}, K.~{Belczynski}, O.~{Korobkin},
  M.~{Chru{\'s}li{\'n}ska}, N.~{Vassh}, M.R. {Mumpower}, J.~{Lippuner}, T.M.
  {Sprouse}, R.~{Surman} et~al., \apj \textbf{855}, 99 (2018),
  \texttt{1710.05875}

\bibitem{vandeVoort2015}
F.~{van de Voort}, E.~{Quataert}, P.F. {Hopkins}, D.~{Kere{\v s}}, C.A.
  {Faucher-Gigu{\`e}re}, \mnras \textbf{447}, 140 (2015), \texttt{1407.7039}

\bibitem{Shen2015}
S.~{Shen}, R.J. {Cooke}, E.~{Ramirez-Ruiz}, P.~{Madau}, L.~{Mayer},
  J.~{Guedes}, \apj \textbf{807}, 115 (2015), \texttt{1407.3796}

\bibitem{Cowan2019}
J.J. {Cowan}, C.~{Sneden}, J.E. {Lawler}, A.~{Aprahamian}, M.~{Wiescher},
  K.~{Langanke}, G.~{Mart{\'{\i}}nez-Pinedo}, F.K. {Thielemann}, arXiv e-prints
   (2019), \texttt{1901.01410}

\bibitem{vandeVoort2019}
F.~{van de Voort}, R.~{Pakmor}, R.J.J. {Grand }, V.~{Springel}, F.A.
  {G{\'o}mez}, F.~{Marinacci}, arXiv e-prints arXiv:1907.01557 (2019),
  \texttt{1907.01557}

\bibitem{Cescutti2015}
G.~{Cescutti}, D.~{Romano}, F.~{Matteucci}, C.~{Chiappini}, R.~{Hirschi}, \aap
  \textbf{577}, A139 (2015), \texttt{1503.02954}

\bibitem{Wehmeyer2015}
B.~{Wehmeyer}, M.~{Pignatari}, F.K. {Thielemann}, \mnras \textbf{452}, 1970
  (2015), \texttt{1501.07749}

\bibitem{Ishimaru2015}
Y.~{Ishimaru}, S.~{Wanajo}, N.~{Prantzos}, \apjl \textbf{804}, L35 (2015),
  \texttt{1504.04559}

\bibitem{Hirai2015}
Y.~{Hirai}, Y.~{Ishimaru}, T.R. {Saitoh}, M.S. {Fujii}, J.~{Hidaka},
  T.~{Kajino}, \apj \textbf{814}, 41 (2015), \texttt{1509.08934}

\bibitem{Ojima2018}
T.~{Ojima}, Y.~{Ishimaru}, S.~{Wanajo}, N.~{Prantzos}, P.~{Fran{\c c}ois}, \apj
  \textbf{865}, 87 (2018), \texttt{1808.03390}

\bibitem{Komiya2016}
Y.~{Komiya}, T.~{Shigeyama}, \apj \textbf{830}, 76 (2016), \texttt{1608.01772}

\bibitem{Bonetti2019}
M.~{Bonetti}, A.~{Perego}, M.~{Dotti}, G.~{Cescutti}, arXiv e-prints
  arXiv:1905.12016 (2019), \texttt{1905.12016}

\bibitem{Chruslinska2018}
M.~{Chruslinska}, K.~{Belczynski}, J.~{Klencki}, M.~{Benacquista}, \mnras
  \textbf{474}, 2937 (2018), \texttt{1708.07885}

\bibitem{Safarzadeh2019a}
M.~{Safarzadeh}, E.~{Ramirez-Ruiz}, J.J. {Andrews}, P.~{Macias}, T.~{Fragos},
  E.~{Scannapieco}, \apj \textbf{872}, 105 (2019), \texttt{1810.04176}

\bibitem{Zevin2019}
M.~{Zevin}, K.~{Kremer}, D.M. {Siegel}, S.~{Coughlin}, B.T.H. {Tsang}, C.P.L.
  {Berry}, V.~{Kalogera}, arXiv e-prints arXiv:1906.11299 (2019),
  \texttt{1906.11299}

\bibitem{Savonije1976}
G.J. {Savonije}, R.J. {Takens}, \aap \textbf{47}, 231 (1976)

\bibitem{DeGreve1977}
J.P. {De Greve}, C.~{De Loore}, \apss \textbf{50}, 75 (1977)

\bibitem{Delgado1981}
A.J. {Delgado}, H.C. {Thomas}, \aap \textbf{96}, 142 (1981)

\bibitem{Tauris2013}
T.M. {Tauris}, N.~{Langer}, T.J. {Moriya}, P.~{Podsiadlowski}, S.C. {Yoon},
  S.I. {Blinnikov}, \apjl \textbf{778}, L23 (2013), \texttt{1310.6356}

\bibitem{Vigna-Gomez2018}
A.~{Vigna-G{\'o}mez}, C.J. {Neijssel}, S.~{Stevenson}, J.W. {Barrett},
  K.~{Belczynski}, S.~{Justham}, S.E. {de Mink}, B.~{M{\"u}ller},
  P.~{Podsiadlowski}, M.~{Renzo} et~al., \mnras \textbf{481}, 4009 (2018),
  \texttt{1805.07974}

\bibitem{Sneden1997}
C.~{Sneden}, R.P. {Kraft}, M.D. {Shetrone}, G.H. {Smith}, G.E. {Langer}, C.F.
  {Prosser}, Astron.~J.~\textbf{114}, 1964 (1997)

\bibitem{Roederer2011}
I.U. {Roederer}, \apjl \textbf{732}, L17 (2011), \texttt{1104.5056}

\bibitem{Sobeck2011}
J.S. {Sobeck}, R.P. {Kraft}, C.~{Sneden}, G.W. {Preston}, J.J. {Cowan}, G.H.
  {Smith}, I.B. {Thompson}, S.A. {Shectman}, G.S. {Burley}, Astron.~J.~ \textbf{141},
  175 (2011), \texttt{1103.1008}

\bibitem{Worley2013}
C.C. {Worley}, V.~{Hill}, J.~{Sobeck}, E.~{Carretta}, \aap \textbf{553}, A47
  (2013), \texttt{1302.6122}

\bibitem{Gratton2012}
R.G. {Gratton}, E.~{Carretta}, A.~{Bragaglia}, \aapr \textbf{20}, 50 (2012),
  \texttt{1201.6526}

\bibitem{Bastian2018}
N.~{Bastian}, C.~{Lardo}, \araa \textbf{56}, 83 (2018), \texttt{1712.01286}

\bibitem{Bekki2017}
K.~{Bekki}, T.~{Tsujimoto}, \apj \textbf{844}, 34 (2017), \texttt{1706.01194}

\bibitem{SAGAdatabase}
Stellar Abundances for Galactic Archaeology Database,
  \texttt{http://sagadatabase.jp}

\bibitem{Suda2008}
T.~{Suda}, Y.~{Katsuta}, S.~{Yamada}, T.~{Suwa}, C.~{Ishizuka}, Y.~{Komiya},
  K.~{Sorai}, M.~{Aikawa}, M.Y. {Fujimoto}, Pub. Astron. Soc. Japan
  \textbf{60}, 1159 (2008)

\bibitem{Battistini2016}
C.~{Battistini}, T.~{Bensby}, \aap \textbf{586}, A49 (2016),
  \texttt{1511.00966}

\bibitem{Cote2017b}
B.~{C{\^o}t{\'e}}, K.~{Belczynski}, C.L. {Fryer}, C.~{Ritter}, A.~{Paul},
  B.~{Wehmeyer}, B.W. {O'Shea}, \apj \textbf{836}, 230 (2017),
  \texttt{1610.02405}

\bibitem{Cote2018b}
B.~{C{\^o}t{\'e}}, M.~{Eichler}, A.~{Arcones}, C.J. {Hansen}, P.~{Simonetti},
  A.~{Frebel}, C.L. {Fryer}, M.~{Pignatari}, M.~{Reichert}, K.~{Belczynski}
  et~al., ArXiv e-prints  (2018), \texttt{1809.03525}

\bibitem{Schoenrich2019}
R.~{Sch{\"o}nrich}, D.H. {Weinberg}, arXiv e-prints  (2019),
  \texttt{1901.09938}

\bibitem{MacFadyen1999}
A.I. {MacFadyen}, S.E. {Woosley}, Astrop. J. \textbf{524}, 262 (1999),
  \texttt{arXiv:astro-ph/9810274}

\bibitem{Stanek+06}
K.Z. {Stanek}, O.Y. {Gnedin}, J.F. {Beacom}, A.P. {Gould}, J.A. {Johnson}, J.A.
  {Kollmeier}, M.~{Modjaz}, M.H. {Pinsonneault}, R.~{Pogge}, D.H. {Weinberg},
  \actaa \textbf{56}, 333 (2006), \texttt{astro-ph/0604113}

\bibitem{Perley+16}
D.A. {Perley}, N.R. {Tanvir}, J.~{Hjorth} et~al., \apj \textbf{817}, 8 (2016),
  \texttt{1504.02479}

\bibitem{Wanderman&Piran10}
D.~{Wanderman}, T.~{Piran}, \mnras \textbf{406}, 1944 (2010),
  \texttt{0912.0709}

\bibitem{Goldstein+16}
A.~{Goldstein}, V.~{Connaughton}, M.S. {Briggs}, E.~{Burns}, \apj \textbf{818},
  18 (2016), \texttt{1512.04464}

\bibitem{Palmerio2019}
J.T. {Palmerio}, S.D. {Vergani}, R.~{Salvaterra}, R.L. {Sanders}, J.~{Japelj},
  A.~{Vidal-Garc{\'\i}a}, P.~{D'Avanzo}, D.~{Corre}, D.A. {Perley}, A.E.
  {Shapley} et~al., arXiv e-prints arXiv:1901.02457 (2019), \texttt{1901.02457}

\bibitem{Bloom2002}
J.S. {Bloom}, S.R. {Kulkarni}, S.G. {Djorgovski}, Astron.~J.~\textbf{123}, 1111
  (2002), \texttt{astro-ph/0010176}

\bibitem{Fruchter2006}
A.S. {Fruchter}, A.J. {Levan}, L.~{Strolger}, P.M. {Vreeswijk}, S.E.
  {Thorsett}, D.~{Bersier}, I.~{Burud}, J.M. {Castro Cer{\'o}n}, A.J.
  {Castro-Tirado}, C.~{Conselice} et~al., \nat \textbf{441}, 463 (2006),
  \texttt{astro-ph/0603537}

\bibitem{Macias2019}
P.~{Macias}, E.~{Ramirez-Ruiz}, \apjl \textbf{877}, L24 (2019),
  \texttt{1905.04315}

\bibitem{Sneden2003}
C.~{Sneden}, J.J. {Cowan}, J.E. {Lawler}, I.I. {Ivans}, S.~{Burles}, T.C.
  {Beers}, F.~{Primas}, V.~{Hill}, J.W. {Truran}, G.M. {Fuller} et~al., \apj
  \textbf{591}, 936 (2003), \texttt{arXiv:astro-ph/0303542}

\bibitem{Winteler2012}
C.~{Winteler}, R.~{K{\"a}ppeli}, A.~{Perego}, A.~{Arcones}, N.~{Vasset},
  N.~{Nishimura}, M.~{Liebend{\"o}rfer}, F.K. {Thielemann}, \apjl \textbf{750},
  L22 (2012), \texttt{1203.0616}

\bibitem{Thompson+04}
T.A. {Thompson}, P.~{Chang}, E.~{Quataert}, Astrophys. J. \textbf{611}, 380
  (2004), \texttt{astro-ph/0401555}

\bibitem{Metzger+08}
B.D. {Metzger}, T.A. {Thompson}, E.~{Quataert}, \apj \textbf{676}, 1130-1150
  (2008), \texttt{0708.3395}

\bibitem{Nishimura2015}
N.~{Nishimura}, T.~{Takiwaki}, F.K. {Thielemann}, \apj \textbf{810}, 109
  (2015), \texttt{1501.06567}

\bibitem{Nishimura2017}
N.~{Nishimura}, H.~{Sawai}, T.~{Takiwaki}, S.~{Yamada}, F.K. {Thielemann}, \apj
  \textbf{836}, L21 (2017), \texttt{1611.02280}

\bibitem{Moesta2014b}
P.~{M{\"o}sta}, S.~{Richers}, C.D. {Ott}, R.~{Haas}, A.L. {Piro},
  K.~{Boydstun}, E.~{Abdikamalov}, C.~{Reisswig}, E.~{Schnetter}, \apjl
  \textbf{785}, L29 (2014), \texttt{1403.1230}

\bibitem{Moesta2018}
P.~{M{\"o}sta}, L.F. {Roberts}, G.~{Halevi}, C.D. {Ott}, J.~{Lippuner},
  R.~{Haas}, E.~{Schnetter}, \apj \textbf{864}, 171 (2018), \texttt{1712.09370}

\bibitem{Halevi&Mosta18}
G.~{Halevi}, P.~{M{\"o}sta}, \mnras  (2018), \texttt{1801.08943}

\bibitem{Sneden2000}
C.~{Sneden}, J.J. {Cowan}, I.I. {Ivans}, G.M. {Fuller}, S.~{Burles}, T.C.
  {Beers}, J.E. {Lawler}, \apjl \textbf{533}, L139 (2000),
  \texttt{astro-ph/0003086}

\bibitem{Frebel2018}
A.~{Frebel}, Ann. Rev. Nucl. Part. Sci. \textbf{68}, 237 (2018),
  \texttt{1806.08955}

\bibitem{Travaglio2004}
C.~{Travaglio}, R.~{Gallino}, E.~{Arnone}, J.~{Cowan}, F.~{Jordan},
  C.~{Sneden}, \apj \textbf{601}, 864 (2004), \texttt{astro-ph/0310189}

\bibitem{Kratz2007}
K.L. {Kratz}, K.~{Farouqi}, B.~{Pfeiffer}, J.W. {Truran}, C.~{Sneden}, J.J.
  {Cowan}, \apj \textbf{662}, 39 (2007), \texttt{astro-ph/0703091}

\bibitem{Siqueira2014}
C.~{Siqueira Mello}, V.~{Hill}, B.~{Barbuy}, M.~{Spite}, F.~{Spite}, T.C.
  {Beers}, E.~{Caffau}, P.~{Bonifacio}, R.~{Cayrel}, P.~{Fran{\c{c}}ois}
  et~al., \aap \textbf{565}, A93 (2014), \texttt{1404.0234}

\bibitem{Winteler+12}
C.~{Winteler}, R.~{K{\"a}ppeli}, A.~{Perego}, A.~{Arcones}, N.~{Vasset},
  N.~{Nishimura}, M.~{Liebend{\"o}rfer}, F.K. {Thielemann}, Astrophys. J. Lett.
  \textbf{750}, L22 (2012), \texttt{1203.0616}

\bibitem{Ji2019}
A.P. {Ji}, M.R. {Drout}, T.T. {Hansen}, arXiv e-prints arXiv:1905.01814 (2019),
  \texttt{1905.01814}

\bibitem{Rosswog+14}
S.~{Rosswog}, O.~{Korobkin}, A.~{Arcones}, F.K. {Thielemann}, T.~{Piran}, Mon.
  Not. R. Astron. Soc. \textbf{439}, 744 (2014), \texttt{1307.2939}

\bibitem{Piro2018}
A.L. {Piro}, J.A. {Kollmeier}, \apj \textbf{855}, 103 (2018),
  \texttt{1710.05822}

\bibitem{Margalit&Metzger17}
B.~{Margalit}, B.D. {Metzger}, \apjl \textbf{850}, L19 (2017),
  \texttt{1710.05938}

\bibitem{Kiziltan2013}
B.~{Kiziltan}, A.~{Kottas}, M.~{De Yoreo}, S.E. {Thorsett}, \apj \textbf{778},
  66 (2013), \texttt{1011.4291}

\bibitem{Gottlieb+18}
O.~{Gottlieb}, E.~{Nakar}, T.~{Piran}, \mnras \textbf{473}, 576 (2018),
  \texttt{1705.10797}

\bibitem{Holmbeck2019}
E.M. {Holmbeck}, T.M. {Sprouse}, M.R. {Mumpower}, N.~{Vassh}, R.~{Surman}, T.C.
  {Beers}, T.~{Kawano}, \apj \textbf{870}, 23 (2019), \texttt{1807.06662}

\bibitem{Schatz2002}
H.~{Schatz}, R.~{Toenjes}, B.~{Pfeiffer}, T.C. {Beers}, J.J. {Cowan},
  V.~{Hill}, K.L. {Kratz}, \apj \textbf{579}, 626 (2002),
  \texttt{astro-ph/0104335}

\bibitem{Roederer2009}
I.U. {Roederer}, K.L. {Kratz}, A.~{Frebel}, N.~{Christlieb}, B.~{Pfeiffer},
  J.J. {Cowan}, C.~{Sneden}, \apj \textbf{698}, 1963 (2009), \texttt{0904.3105}

\bibitem{Mashonkina2014}
L.~{Mashonkina}, N.~{Christlieb}, K.~{Eriksson}, \aap \textbf{569}, A43 (2014),
  \texttt{1407.5379}

\bibitem{Metzger2014b}
B.D. {Metzger}, A.L. {Piro}, \mnras \textbf{439}, 3916 (2014),
  \texttt{1311.1519}

\bibitem{Siegel2016a}
D.M. {Siegel}, R.~{Ciolfi}, \apj \textbf{819}, 14 (2016), \texttt{1508.07911}

\bibitem{Siegel2016b}
D.M. {Siegel}, R.~{Ciolfi}, \apj \textbf{819}, 15 (2016), \texttt{1508.07939}

\end{thebibliography}

\end{document}